\documentclass[trackchanges,twocolumn]{aastex701} 
\usepackage{amsmath}
\usepackage{booktabs}
\usepackage{multirow}
\usepackage{threeparttable}
\usepackage{makecell}
\usepackage{soul}
\usepackage{graphicx}
\usepackage{subcaption}
\usepackage{float}
\usepackage[normalem]{ulem}
\usepackage{cancel}
\usepackage{soul}

\usepackage{tabu}
\usepackage{ltablex}
\usepackage{array}
\usepackage{hyperref}
\usepackage{comment}

\begin{document}

\title{Co-evolution of AGN and Host Galaxies across Cosmic Time Using JWST data}
\author[0000-0002-0524-5328]{Priyanka Jalan}
\affiliation{Institute of Astronomy, National Tsing Hua University, 101, Section 2, Kuang-Fu Road, Hsinchu, 30013, Taiwan}
\affiliation{Center for Theoretical Physics, Polish Academy of Sciences, Al. Lotnik\'ow 32/46, 02-668, Warsaw, Poland}
\email[show]{priyajalan14@gmail.com}
\correspondingauthor{Priyanka Jalan}

\author[0000-0002-6821-8669]{Tomotsugu Goto}
\affiliation{Institute of Astronomy, National Tsing Hua University, 101, Section 2, Kuang-Fu Road, Hsinchu, 30013, Taiwan}
\affiliation{Department of Physics, National Tsing Hua University, 101, Section 2. Kuang-Fu Road, Hsinchu, 30013, Taiwan}
\email[]{itlhnis@gmail.com}

\author[0000-0001-7228-1428]{Tetsuya Hashimoto}
\affiliation{Department of Physics, National Chung Hsing University, 145, Xingda Road, Taichung, 40227, Taiwan}
\email[]{tetsuya@phys.nchu.edu.tw }

\author[0009-0009-6588-2271]{Angel Rodriguez Barbosa}
\affiliation{Institute of Astronomy, National Tsing Hua University, 101, Section 2, Kuang-Fu Road, Hsinchu, 30013, Taiwan}
\email[]{angelskyship@gmail.com }

\author[0000-0002-3738-1834]{Chih-Teng Ling}
\affiliation{Department of Astronomical Science, SOKENDAI (The Graduate University for Advanced Studies), 2-21-1 Osawa, Mitaka, Tokyo, 181-8588, Japan}
\affiliation{National Astronomical Observatory of Japan, 2-21-1 Osawa, Mitaka, Tokyo 181-8588, Japan}
\email[]{chih-teng.ling@grad.nao.ac.jp}

\author[0000-0002-9119-2313]{Ece Kilerci}
\affiliation{Department of Astronomy and Space Sciences, Istanbul University, Beyazıt, 34452 Fatih/İstanbul, Türkiye}
\email[]{ecekilerci@gmail.com }

\author[0009-0009-2940-0861]{Terry Long Phan}
\affiliation{Institute of Astronomy, National Tsing Hua University, 101, Section 2, Kuang-Fu Road, Hsinchu, 30013, Taiwan}
\email[]{phanngoclong224@gmail.com }

\author[]{Srivardini Ayyappan }
\affiliation{Institute of Astronomy, National Tsing Hua University, 101, Section 2, Kuang-Fu Road, Hsinchu, 30013, Taiwan}
\email[]{srivardini.ayyappan@hdr.mq.edu.au }

\begin{abstract}

Understanding the complex co-evolution of active galactic nuclei (AGN) and their host galaxies across cosmic time remains a fundamental challenge, particularly in identifying how the physical drivers of galaxy diversity evolve from the early Universe to the present day. Constraining these relationships is critical for establishing a unified framework of galaxy evolution that connects black hole growth and star formation. To understand this, we use a multi-wavelength JWST catalog of 2,709 sources cross-matched from the SMILES and JWST Advanced Deep Extragalactic Survey (JADES) surveys, spanning lookback times from 0.8 to 12.9 Gyr ($0.05 <z <6$).  Using Pearson correlation matrix, we find AGN luminosity ($L_{\mathrm{AGN}}$) strongly correlates ($\sim0.8$) with  star formation rate (SFR) and dust luminosity  ($L_{\mathrm{dust}}$), weakly correlates ($\sim0.5$) with stellar mass ($M_\star$) and anti-correlates  ($\sim -0.4$) with star formation history (SFH) duration and stellar age (Age$_\star$). However, the apparent anti-correlation observed between galaxy age and AGN luminosity is driven by cosmic time rather than by a direct physical connection.

Using Principal Component Analysis (PCA), we find that galaxy diversity is fundamentally governed by two principal components: AGN and galaxy activity, and stellar age. At higher redshifts, AGN accretion, star formation, and dust emission are tightly synchronized, reflecting gas-rich conditions in the early Universe. In contrast, the present Universe exhibits a clear bimodal structure in host-dominated systems ($f_{\mathrm{AGN}} \leq 0.2$), separating quiescent and star-forming galaxies. Our results demonstrate that AGN luminosity is primarily governed by the host galaxy's current cold-gas supply, of which recent star formation is a close observational tracer, with galaxy mass setting the overall potential.
This highlights a co-evolution scenario in which black hole growth and star formation are tightly linked, particularly at early cosmic epochs.

\end{abstract}


\section{Introduction}

The co-evolution of supermassive black holes and their host galaxies remains a central problem in modern astrophysics \citep{Kormendy2013, Heckman2014, Yang2018, Kim_Im_Jun_Kim_2021, Mountrichas2023, Capelo2024}. Observational scaling relations, such as those between black hole mass and stellar bulge properties, suggest a deep physical connection between galaxy growth and black hole accretion \citep{Magorrian1998, Ferrarese2000, Gebhardt2000}. Active Galactic Nuclei (AGN) represent a key phase in this connection, during which accretion onto a central supermassive black hole (SMBH) releases vast amounts of energy capable of regulating star formation, heating interstellar gas, and driving galaxy-scale feedback \citep{Silk1998, Fabian2012, Ishibashi2012, King2015}. Understanding how AGN energy output relates to global galaxy properties-such as stellar mass, dust content, and star formation history-is therefore essential for tracing the buildup of structure across cosmic time \citep{Merloni2010, Aird2012, Aversa2015, Ciesla2015, Suh2019, Stemo_2020,  Barrows_2021, Mountrichas2023a, Sun2025, Terrazas2025}.

A fundamental open question is what physical mechanisms primarily govern the accretion rate onto supermassive black holes, and thus regulate AGN luminosity, and how these processes evolve with redshift 
\citep{Hopkins2006, Alexander2012, Shen2020, Alexander2025, Li2026}. One of the scenario proposed for triggering or enhancing AGN activity is galaxy interactions and mergers through merger-driven gas inflows \citep{Sanders1988, Hopkins2006, Ellison2011}, but observational evidence for a universal merger-AGN connection remains mixed \citep{Cisternas2011, Kocevski2012, Mechtley2016, Villforth2017, Blecha2018, Ellison2019, Omori2025}. While some studies find little or no merger excess in AGN hosts \citep{Grogin2005, Li2008, Gabor2009, Schawinski2012, Sabater2015, Marian2019, Villforth2023, Sharma2024}, others find enhanced AGN activity in interacting systems \citep{ Silverman2011, Koss2012, Satyapal2014, Goulding2018, Gao2020, Bonaventura2026, Marca2026}. Although mergers may play an important role in triggering some AGN, however with the complexity of merger depending on AGN luminosity, merger stage, redshift, with secular processes, the availability of cold gas \citep{Mechtley2016, Blecha2018, Omori2025}, and lack of direct merger diagnostics in our analysis, we do not investigate merger signatures here and instead focus on the intrinsic properties of the host galaxies to determine which physical properties are most closely associated with black hole accretion.

In context of physical properties, it also remains unclear whether AGN power is governed predominantly by the integrated properties of the host galaxy, such as its total stellar mass and long-term assembly history, or by more immediate conditions, including the availability of cold gas and ongoing star formation \citep{Mullaney2012, Rosario2012, Chen2013}. This distinction is crucial for constraining theoretical models of galaxy evolution and black hole growth. 

Several studies have reported a weak or nearly flat relation with these intrinsic properties after accounting for selection effects and AGN variability, suggesting that the short-term variability of AGN accretion can obscure any underlying long-term connection with galaxy growth \citep{Ma2015, Stanley2015, Pitchford2016, Shimizu2017}. Conversely, other works have found that luminous AGN exhibit enhanced star formation, particularly at high AGN luminosities or within specific redshift ranges, implying that both black hole accretion and star formation may be regulated by a common cold gas reservoir \citep{Shao2010, Rosario2012, Harris2016}. Stellar mass is generally expected to correlate positively with AGN luminosity through the established scaling relation between galaxy stellar mass and black hole mass, although its importance appears to diminish at higher redshifts where the SFR becomes the dominant predictor of black hole accretion \citep{Delvecchio2015}. In contrast, stellar population ages are expected to show, at most, weak correlations with AGN luminosity because they trace the integrated star formation history rather than the instantaneous gas supply that fuels accretion. Likewise, previous studies have found little or no dependence of AGN activity on black hole Eddington ratio, X-ray obscuration, or other host properties once selection effects are considered \citep{Shao2010, Rosario2012, Harris2016, Pitchford2016}. A robust assessment of these relationships therefore requires controlling for cosmic evolution and luminosity-dependent selection effects by examining the correlations within narrow redshift intervals, rather than relying solely on the trends observed in the full sample.

Addressing this problem is challenging for several reasons. AGN activity is inherently episodic and operates on relatively short timescales compared to the billion-year timescales over which galaxies assemble their stellar mass \citep{Hickox2014, Schawinski2015}. In addition, the observational separation of AGN emission from star formation is non-trivial, as both processes contribute significantly to the infrared and ultraviolet emission of galaxies, particularly in dust-rich systems \citep{Netzer2009, Kirkpatrick2015}. These degeneracies have historically limited the ability to isolate the physical drivers of AGN activity. Furthermore, until recently, the lack of deep, continuous multi-wavelength data at high redshift has hindered efforts to determine whether AGN accretion and star formation were more tightly coupled in the early universe \citep{Madau2014, Casey2014} or not.

In this work, we address these challenges using a comprehensive multi-wavelength dataset of 2,709 galaxies spanning lookback times from 0.8 to 12.9 Gyr ($0.05 < z < $6). The dataset combines optical, near-infrared, and mid-infrared observations, enabling robust spectral energy distribution (SED) fitting and the simultaneous estimation of key physical parameters. These include stellar mass ($M_*$), instantaneous star formation rates ($\mathrm{SFR}_{10\,\mathrm{Myr}}$), dust luminosity ($L_{\rm dust}$), stellar age ($\mathrm{Age}_*$), and AGN bolometric luminosity ($L_{\rm AGN}$; integrated over the AGN SED components), along with structural properties such as effective radius.

To isolate the dominant drivers of AGN luminosity, we adopt a two-fold analytical approach. First, we perform correlation analysis to quantify the relationships between AGN power and host galaxy properties. Second, we apply principal component analysis \citep[PCA;][]{Boroson1992, Francis1992, Corbin1996, Jin2012, Davies2018} to reduce the dimensionality of the dataset and identify the primary axes of variance, allowing us to track how galaxy populations organize in an ``activity-evolution'' parameter space across cosmic time. 

Our results show that AGN luminosity is most strongly correlated with indicators of recent activity, particularly star formation rate and dust luminosity, consistent with a shared dependence on cold gas reservoirs \citep{Bongiorno2012, Azadi2015, Trump2015, Mountrichas2023}.  In contrast, weaker correlations with structural parameters suggest that integrated galaxy properties play a secondary role. PCA further reveals that the majority of variance ($>90\%$) is captured by two components: an activity (or scale) axis linking AGN power, star formation, and dust emission, and an evolutionary axis associated with stellar age.

The paper is organized as follows: In Section~\ref{sec:Data}, we describe the multi-wavelength data used in this study. In Section~\ref{sec:corr}, we make the module-wise correlation analysis and find the parameters correlated with AGN luminosity. We then study the correlated parameters versus the AGN luminosity in Section~\ref{sec:evol}, dividing the sample based on the AGN fraction to host dominated and AGN dominated and investigate their redshift evolution. In Section~\ref{sec:pca}, we perform a PCA to study the combined effects of the correlated physical parameters on AGN activity, further subdividing the sample into different lookback-time bins to trace evolutionary trends. Finally, we discuss the findings in Section~\ref{sec:conc}. Throughout, we adopt a flat $\Lambda$CDM cosmology consistent with \citet{Planck2020params}  unless otherwise specified.

\section{Data}
\label{sec:Data}

This study uses the multi-wavelength galaxy catalog presented by \citet{Ling2026}, which combines imaging from two major \textit{James Webb Space Telescope (JWST)} legacy programs: the \textit{Systematic Mid-infrared Instrument Legacy Extragalactic Survey (SMILES)} and the \textit{JWST Advanced Deep Extragalactic Survey (JADES)}. Together, these surveys provide homogeneous photometry spanning observed-frame wavelengths from 0.4 to 25.5~$\mu$m across the GOODS-S field. The broad wavelength coverage enables simultaneous modeling of stellar populations, dust emission, and AGN-heated dust using spectral energy distribution (SED) fitting. Since the observed wavelength coverage extends only to 25.5~$\mu$m, the rest-frame mid-infrared regime that most strongly constrains obscured AGN becomes progressively less accessible at the highest redshifts. Consequently, in this work we restrict our analysis to galaxies with $z<6$, avoiding the most extreme systems where the AGN contribution becomes poorly constrained.

A detailed description of the catalog is presented in \citet{Ling2026} and \citet{Barbosa_2026}; we summarize the main aspects below. SMILES \citep[PID 1207;][]{SMILES_Rieke2024} provides continuous imaging with JWST/MIRI in eight filters (F560W, F770W, F1000W, F1280W, F1500W, F1800W, F2100W, and F2550W), covering 5.6--25.5~$\mu$m over approximately 34.5 arcmin$^2$ in GOODS-S/HUDF. Exposure times range from 650--2100~s per band. The multiple MIRI bands substantially reduce degeneracies between dust heated by star formation and AGN-heated dust compared to near-infrared photometry alone, although these constraints naturally weaken with increasing redshift as progressively shorter rest-frame wavelengths are sampled.

JADES \citep[PID 1180;][]{Eisenstein2026} provides deep NIRCam imaging in nine broadband filters (F090W--F444W), complemented by five medium-band filters from the JEMS survey \citep[PID 1963;][]{JEMS_2023ApJS..268...64W}, together with nine HST/WFC3 and ACS bands from the CANDELS survey \citep{CANDELS_2011ApJS..197...35G,CANDELS_2011ApJS..197...36K}. The combined optical and near-infrared dataset comprises 23 bands.

The SMILES and JADES catalogs were cross-matched using a matching radius of 0.2\arcsec\ \citep{JADES_Rieke2023ApJS..269...16R,SMILES_Alberts2024}, producing a multi-wavelength catalog. Source detection was performed on the stacked F560W+F770W image, with aperture photometry measured in all MIRI bands. The astrometric accuracy is 0.1--0.2\arcsec\ for F560W--F2100W and approximately 0.4\arcsec\ for F2550W.

Photometric redshifts were derived using \texttt{EAZY} \citep{Brammer2008ApJ...686.1503B}, employing linear combinations of the seven templates from \citet{Hainline2024ApJ...964...71H} together with the nine default \texttt{EAZY} templates, Kron photometry, and a 5\% systematic uncertainty. Validation against 276 spectroscopic redshifts obtained with NIRSpec demonstrated that the NIR-only photometric redshifts outperform those obtained using combined NIR+MIR photometry ($\sigma_{\rm NMAD}=0.0169$, $\eta=3.6\%$ versus $\sigma_{\rm NMAD}=0.0194$, $\eta=6.2\%$), owing to larger uncertainties in the MIRI photometry and the limited availability of MIR templates. Accordingly, the NIR-only photometric redshifts (supplemented by spectroscopic redshifts where available) were adopted. These redshifts were subsequently fixed during the derivation of physical properties by \texttt{CIGALE} \citep{Boquien2019A&A...622A.103B}, substantially improving computational efficiency while preserving the full SED modeling capability.  The fixed-redshift approach is particularly appropriate here because our objective is to derive robust galaxy physical properties rather than to jointly optimize the redshift and SED parameters.

\texttt{CIGALE} v2022.1 fitting uses the complete 31-band photometry (23 optical/near-IR + 8 MIRI bands). The models include delayed star formation histories ($\text{SFR} \propto t \exp[-t/\tau]$; characterized by a stellar population age, $t$, and an e-folding decay time, $\tau$), \citet{Chabrier2003PASP..115..763C} initial mass functions with BC03 stellar populations, dust attenuation \citep[\texttt{dustatt\_modified\_CF00;}][]{Charlot2000ApJ...539..718C}, dust emission \citep[\texttt{dl2014;}][]{Draine2014ApJ...780..172D}, and AGN emission represented by the \texttt{SKIRTOR2016} \citep{Stalevski2016MNRAS.458.2288S} torus models. 
Following the parameter configuration adopted in \citet[][that adopts parameters from \citet{Yang2023ApJ...950L...5Y}]{Ling2026}, the explored AGN parameter space preferentially selects obscured (Type$\sim$2-like) AGN solutions.

For clarity, we define here the physical quantities used throughout this work. 
$L_{\mathrm{AGN}}$ is the total AGN luminosity output by \texttt{CIGALE} using the \texttt{SKIRTOR2016} AGN module \citep{Stalevski2016MNRAS.458.2288S}, integrated over the full modeled AGN SED, with a fixed inclination corresponding to an edge-on, type-2-like solution. $L_{\mathrm{dust}}$ is the infrared luminosity of the star-formation-heated dust component, computed by the \texttt{dl2014} module \citep{Draine2014ApJ...780..172D} via energy balance with the dust-attenuated stellar continuum.

The \texttt{dl2014} module additionally provides several parameters used in our correlation analysis (Figure~\ref{fig:corrdust}): $q_{\mathrm{PAH}}$, the mass fraction of the dust grain population in polycyclic aromatic hydrocarbons (PAHs), sampled on a grid of $\{0.47, 2.5, 7.32\}\%$; $U_{\mathrm{min}}$, the minimum interstellar radiation field intensity (in units of the local Galactic value) heating the diffuse dust component, sampled at $\{10, 15, 20\}$; and $\Gamma_{\mathrm{dust}}$, the fraction of the dust mass exposed to a power-law distribution of radiation field intensities above $U_{\mathrm{min}}$ (i.e., dust associated with photo-dissociation regions), sampled at $\{0.01, 0.02, 0.05, 0.1, 0.2, 0.5\}$. $U_{\mathrm{mean}}$ is not an independent fit parameter but is derived from $U_{\mathrm{min}}$ and $\Gamma_{\mathrm{dust}}$ following the standard \texttt{dl2014} formalism, and represents the mass-weighted mean radiation field intensity across the full dust distribution.

The stellar mass ($M_\star$) and the associated gas mass ($M_{\mathrm{gas}}$) are derived within the BC03 stellar-population framework from the fitted star-formation history and its normalization. Here, $M_{\mathrm{gas}}$ represents the gas mass associated with stellar evolution and mass return in the stellar-population model. The BC03 module further separates these quantities into young and old populations using a separation age of 10 Myr: $M_\star^{\mathrm{young}}$ and $M_{\mathrm{gas}}^{\mathrm{young}}$ correspond to the young population, while $M_\star^{\mathrm{old}}$ and $M_{\mathrm{gas}}^{\mathrm{old}}$ correspond to the older population.

The AGN contribution is inferred through full UV-to-mid-infrared SED decomposition rather than from any individual photometric band. The eight MIRI filters provide significantly stronger constraints on the AGN component than analyses relying on only one or two mid-infrared measurements. Nevertheless, we note that the uncertainty in the inferred AGN fraction increases toward higher redshift as the characteristic hot-dust emission shifts beyond the observed wavelength range. Consequently, the derived AGN fractions should be interpreted within the limitations imposed by the available rest-frame wavelength coverage.

Following previous SMILES studies, we characterize galaxies using the infrared AGN fraction,
\[
f_{\rm AGN}=\frac{L_{\rm AGN}}{L_{\rm TIR}},
\]
where $L_{\rm AGN}$ and $L_{\rm TIR}$ denote the AGN infrared luminosity and total infrared luminosity (representing the sum of the dust-reprocessed emission from both star formation and the active galactic nucleus), respectively, defined over a restricted rest-frame range of 3-30 $\mu$m following \citet[][following \citealt{Yang2023ApJ...950L...5Y}]{Ling2026}, and $f_{\rm AGN}$ is used solely for the host/composite/AGN classification in further sections. Systems are classified as star-forming galaxies ($f_{\rm AGN}<0.2$), composites ($0.2\le f_{\rm AGN}<0.5$), and AGNs ($f_{\rm AGN}\ge0.5$).

Consequently, in this work we restrict our analysis to galaxies with $z<6$, avoiding the most extreme systems where the AGN contribution becomes poorly constrained. The final sample consists of 2709 galaxies. Among these, 2431 are classified as host-dominated ($f_{\rm AGN}\le0.2$), while the remaining 278 comprise composite and AGN systems ($f_{\rm AGN}>0.2$). Throughout this work we combine the composite and AGN populations into a single AGN-dominated sample.

This curated multi-wavelength catalog forms the basis of our investigation of the connection between AGN activity and host-galaxy properties over the redshift range $0<z<6$. Physical quantities including stellar mass, gas mass, star formation rate, and AGN luminosity are analyzed in logarithmic space with propagated uncertainties. Parameters that are fixed during the fitting procedure or remain poorly constrained (e.g., torus geometry or inclination when effectively fixed by the adopted model grid) are excluded from the statistical analysis to minimize degeneracies and avoid introducing redundant information.

\section{Correlation Analysis}
\label{sec:corr}
For each module from \texttt{CIGALE} fitting (AGN, stellar, dust, attenuation, and structural), we compute the Pearson correlation matrix (Figure.~\ref{fig:corrsfh}-\ref{fig:corrdust}). Parameters that are highly correlated with the AGN luminosity ($L_{\rm AGN}$) above a threshold of $r_{\rm min}=0.4$ are selected for further analysis. We adopt this threshold to minimize the inclusion of weak or noise-dominated relationships while preserving variables that trace physically meaningful connections with AGN luminosity.

To test the robustness in the observed correlation, we propagate the measurement uncertainties of the \texttt{CIGALE} parameters into the correlation coefficients, using Monte Carlo uncertainty analysis. For each realization, every parameter is perturbed according to a Gaussian distribution centered on its measured value with a standard deviation equal to its reported $1\sigma$ uncertainty. A new Pearson correlation matrix is then computed, and this procedure is repeated 1000 times. The uncertainty is estimated from the central 68\% confidence interval,
\begin{equation}
\sigma_r = \frac{r_{84}-r_{16}}{2},
\end{equation}
where $r_{16}$ and $r_{84}$ are the 16th and 84th percentiles of the Monte Carlo distribution of the correlation coefficient. Error bars are shown only when the uncertainty is $\geq 0.01$.

Redundant parameters with $|r|\gtrsim0.95$ are subsequently pruned. For a redundant pair $(X_i, X_j)$ satisfying $|r(X_i,X_j)|\gtrsim0.95$, we retain only the parameter that exhibits the stronger correlation with $L_{\rm AGN}$,
\begin{equation}
\mathrm{keep}\; X_i
\quad \mathrm{if} \quad
|r(X_i,L_{\rm AGN})|
>
|r(X_j,L_{\rm AGN})|.
\end{equation}

The resulting set of highlighted parameters (e.g. $\mathrm{SFR}_{10\,\mathrm{Myr}}$, $M_\star$, and SFH Age) shown in Figure.~\ref{fig:corrsfh}--\ref{fig:corrdust} is used for the subsequent analysis. Module-wise grouping is adopted for clarity of visualization and to facilitate the physical interpretation of the relationships with AGN luminosity.

\subsection*{Star Formation History (SFH) Module}
\begin{figure}[ht]
    \centering
    \includegraphics[width=0.5\textwidth]{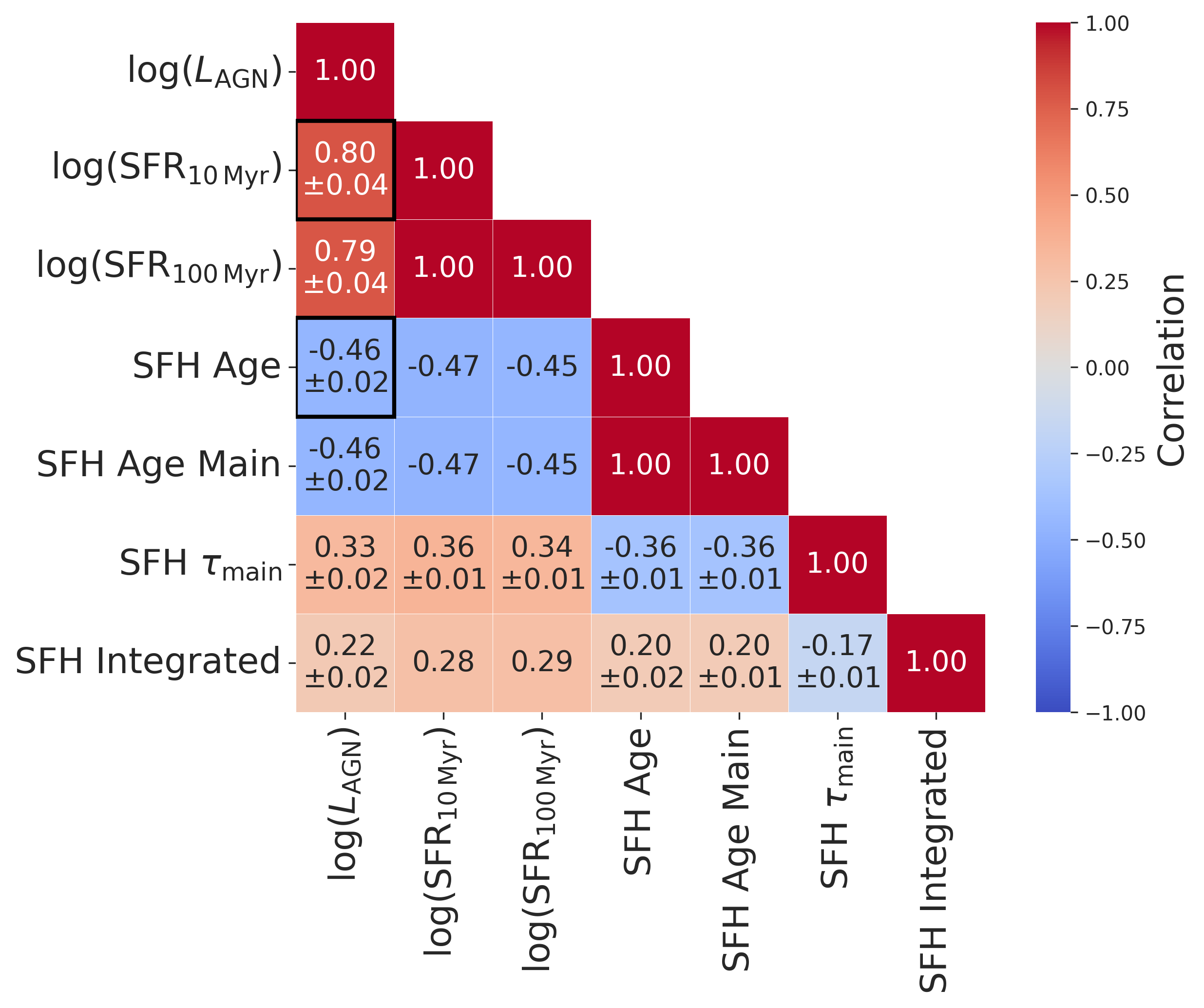}
    \caption{Correlation matrix illustrating the physical relationships between AGN luminosity ($\log L_{\mathrm{AGN}}$), with in the SFH module including star formation rate (SFR), and star formation history (SFH) parameters.}
    \label{fig:corrsfh}
\end{figure}

\begin{figure*}[ht]
    \centering
    \includegraphics[width=0.95\textwidth]{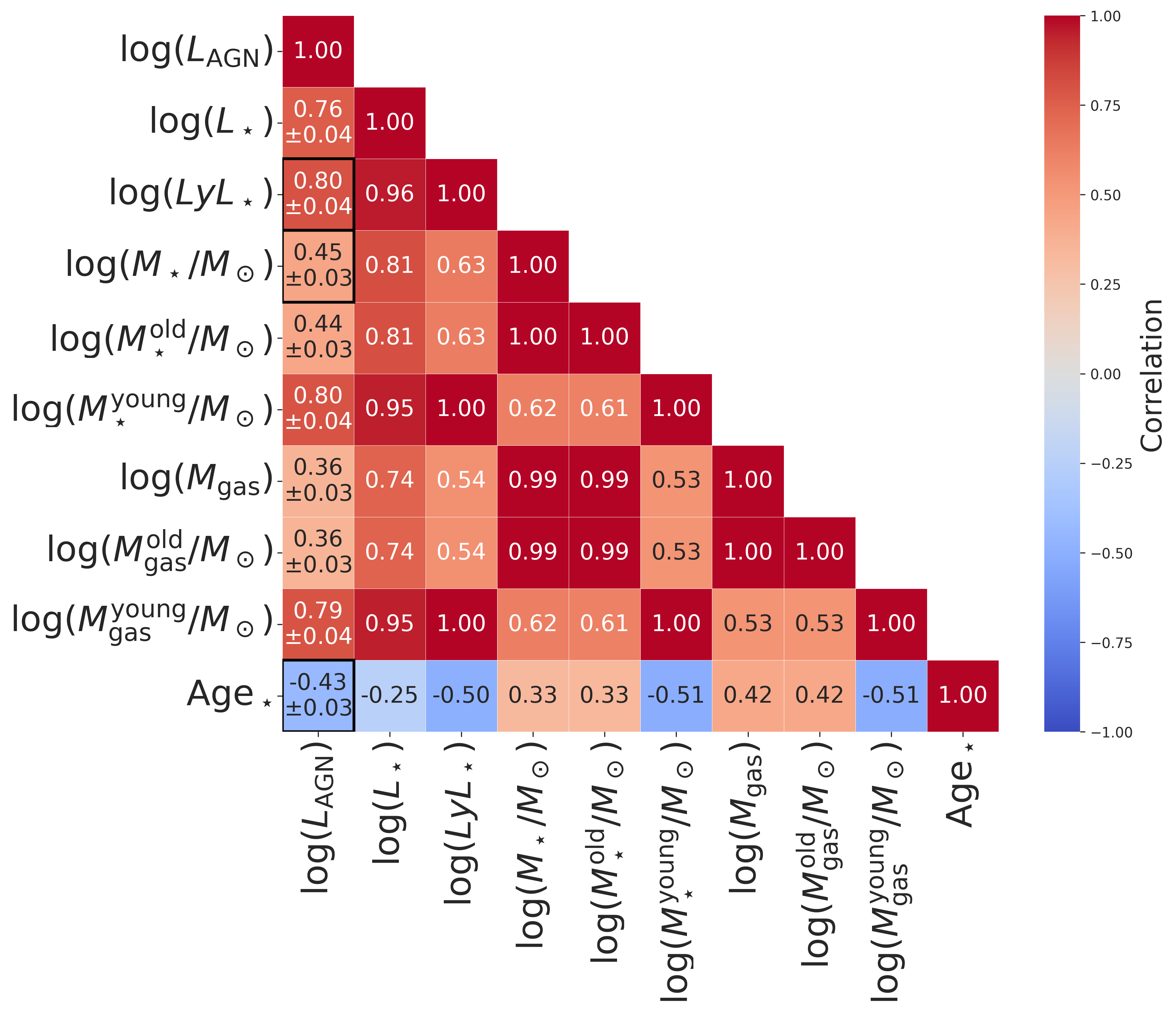}
    \caption{Same as Figure.~\ref{fig:corrsfh} but for stellar module.}
    \label{fig:corrstellar}
\end{figure*}

This module highlights the physical coupling between black hole growth and star formation (Figure.~\ref{fig:corrsfh}):
\begin{itemize}
    \item Strong positive correlation with SFR: $L_{\rm AGN}$ is very strongly linked to recent star formation rates, with a correlation of $0.80$ for $\mathrm{SFR}_{10\,\rm Myr}$ and $0.79$ for $\mathrm{SFR}_{100\,\rm Myr}$.  This suggests that both AGN activity and star formation are fueled by a common reservoir of cold gas. Dense gas inflows that trigger rapid star formation also provide material for accretion onto the central black hole.
    \item Negative correlation with age:  SFH age represents the characteristic age of a galaxy's stellar population derived from its star formation history, typically reflecting the time since the bulk of its stars were formed. We find a negative correlation with SFH Age ($-0.46$), suggesting that more luminous AGNs are typically found in ``younger,'' more actively forming galaxies. As galaxies evolve, their gas reservoirs are progressively depleted or expelled through feedback processes, leading to a decline in both star formation and black hole accretion. In contrast, younger systems are more likely to experience gas inflows, mergers, or dynamical instabilities that channel gas toward the central regions, thereby enhancing AGN activity.
\end{itemize}

\subsection*{Stellar Module}
The stellar parameters reinforce the trends seen in the SFH module  (Figure.~\ref{fig:corrstellar}) in addition to some additional correlations as follows:
\begin{itemize}
    \item Total stellar mass: Based on the correlation matrix and the pruning methodology, $M_\star$ remains selected (black-squared) while $M_\star^{\mathrm{young}}$ is excluded due to redundancy rules. Although $M_\star^{\mathrm{young}}$ shows a much stronger correlation with AGN luminosity ($r = 0.80$), indicating that AGN activity is primarily driven by the availability of cold gas, it is highly redundant ($|r| \sim 1$) with other young-component tracers such as Lyman luminosity  (${\mathrm{Ly}} L_*$)  and $M_{\mathrm{gas}}^{\mathrm{young}}$. Within this redundant group, the parameter selection is based on the strongest correlation with $L_{\mathrm{AGN}}$, and ${\mathrm{Ly}} L_\star$ is retained as the representative ``young activity'' tracer. In contrast, $M_\star$ is part of a different redundant group dominated by older stellar components (e.g. $M_\star^{\mathrm{old}})$ and $(M_{\mathrm{gas}})$. Within this group, it is selected because it has the strongest correlation with AGN luminosity ($|r| = 0.45$) compared to $M_\star^{\mathrm{old}}$ ($r = 0.44$) and $M_{\mathrm{gas}}$ ($r = 0.36$, below threshold). The weaker relation with total stellar mass is therefore a result of dilution by older stellar populations, which dominate the mass budget but do not contribute significantly to ongoing AGN fueling. This demonstrates that AGN luminosity is more closely linked to the galaxy's current growth phase than to its cumulative stellar mass. 
    \item Stellar age ($\mathrm{Age}_*$): Consistent with the SFH module, there is a clear anti-correlation ($-0.43$) with the average stellar age i.e., age of stars/light-dominating population.
    \item Lyman luminosity: Lyman luminosity refers to the luminosity emitted in transitions involving hydrogen's Lyman series, particularly Ly$\alpha$ emission at 1216 \AA, commonly used to trace star formation and ionized gas in galaxies. Based on the correlation matrix, $L_{\mathrm{AGN}}$ and ${\mathrm{Ly}} L_\star$ exhibit a strong positive correlation of $\sim 0.80$. This indicates that the power of the central AGN is closely linked to high-energy radiation associated with young stellar populations and substantial gas reservoirs. However, this parameter becomes redundant when correlated with SFR. 
\end{itemize}

\subsection*{Dust Module}
\begin{figure}[ht]
    \centering
    \includegraphics[width=0.5\textwidth]{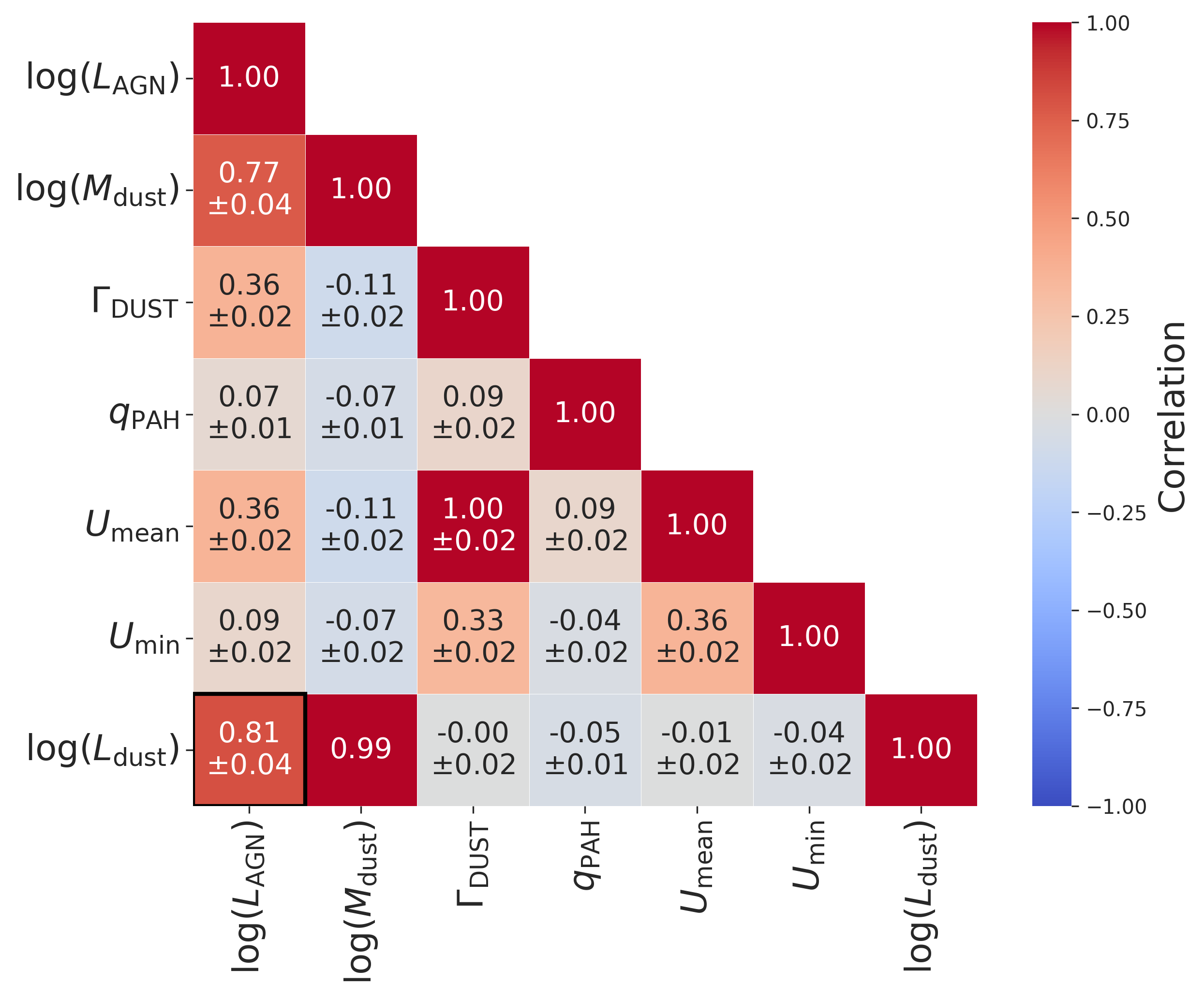}
    \caption{Same as Figure.~\ref{fig:corrsfh} but for dust module.}
    \label{fig:corrdust}
\end{figure}

This module highlights the physical coupling between
black hole growth and the dust properties  (Figure.~\ref{fig:corrdust}):
\begin{itemize}
    \item Dust content: The strong correlations between $L_{\mathrm{AGN}}$,  with dust luminosity ($L_{\mathrm{dust}} \sim 0.81$) and dust mass ($M_{\mathrm{dust}} \sim 0.77$) reflect the close physical coupling between black hole growth and the host galaxy's gas and dust reservoirs. Dust mass traces the cold gas supply available for both star formation and black hole accretion, naturally linking higher $M_{\mathrm{dust}}$ to more luminous AGN activity. 
Dust luminosity arises from the reprocessing of radiation, where energetic 
output from young stars is absorbed and re-emitted in the infrared. Because 
star formation and AGN accretion are both thought to be fueled by a common 
cold-gas reservoir, this star-formation-heated dust emission traces the same 
gas-rich conditions that regulate AGN fueling, producing the observed scaling 
between $L_{\mathrm{dust}}$ and $L_{\mathrm{AGN}}$. These trends are further 
reinforced by the strong connection between AGN luminosity and recent star 
formation, indicating that luminous AGNs preferentially reside in actively 
growing, gas-rich systems. In such environments, stellar mass assembly and 
black hole accretion proceed concurrently. Finally, the near-perfect 
correlation between $L_{\mathrm{dust}}$ and $M_{\mathrm{dust}}$ ($\sim 0.99$) 
suggests mathematically tied parameters, and only $L_{\mathrm{dust}}$ is used 
for further analysis.

\end{itemize}

\begin{figure*}[!ht]
    \centering
    \includegraphics[width=0.99\textwidth]{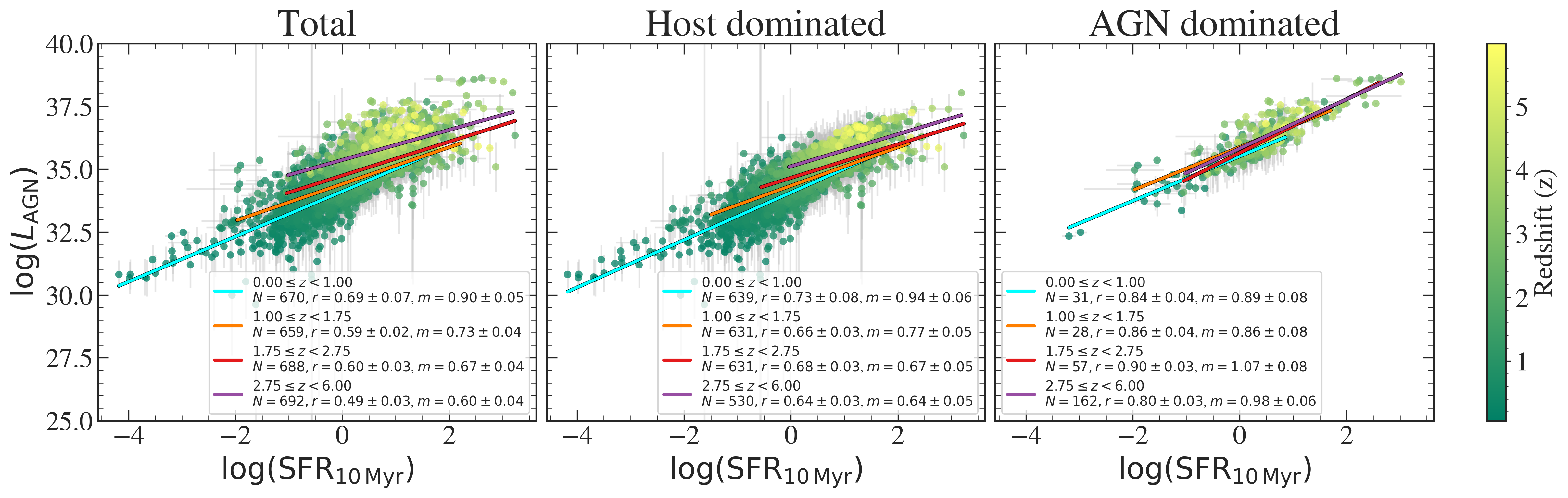}
        \includegraphics[width=0.99\textwidth]{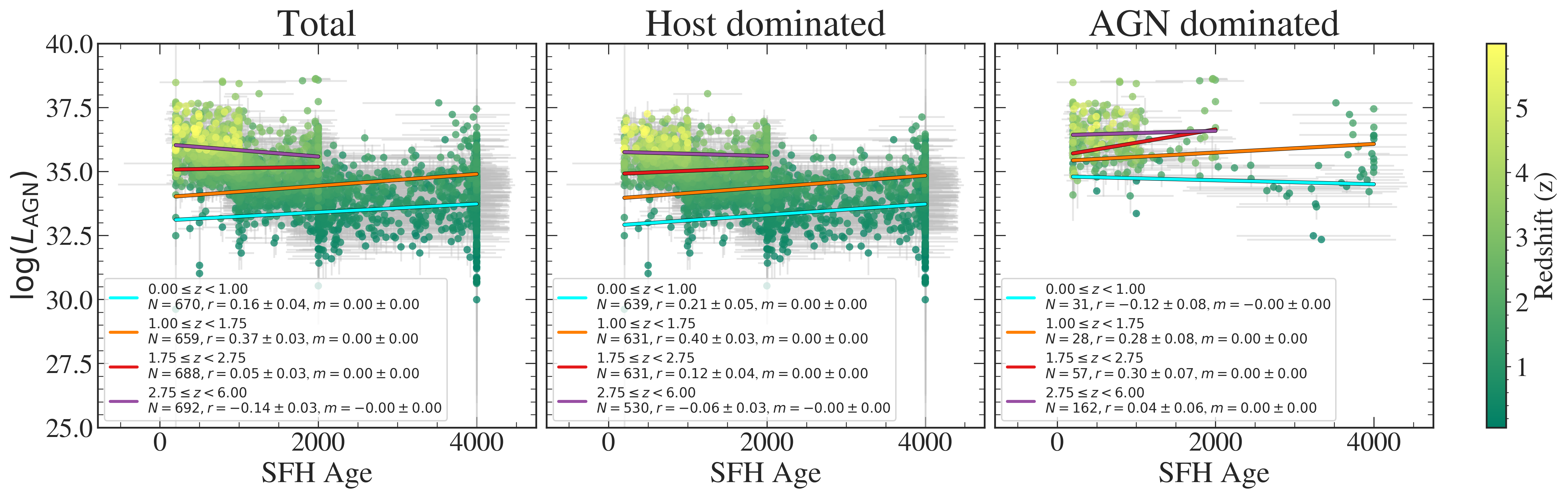}
    \caption{{\it Top panel}: AGN luminosity versus recent star formation rate (${\rm SFR}_{10\,\mathrm{Myr}}$) for total [left], host-dominated [middle], and AGN-dominated samples [right]. The color bar shows the redshift of the sources. {\it Bottom panel}: Correlation between AGN luminosity ($L_{\rm AGN}$) and the duration of the star formation history (SFH Age [Myr]).}
    \label{fig:vssfr}
\end{figure*}

\subsection*{Other noticeable correlation/non-correlations}
\begin{itemize}

\item The effective radius ($R_{\rm e}$) is a global structural parameter that describes the overall size of a galaxy (the radius within which half its light is emitted). The effective radius was computed as the geometric mean of the semi-major and semi-minor axes, $R_{\rm e}=\sqrt{ab}$ and converted from angular units (arcsec) to physical units (kpc) using the proper angular scale at the galaxy redshift. The sources show that while $R_{\rm e}$ has a strong positive correlation with total stellar mass (0.73) and total gas mass (0.75), it has almost no direct relationship with the energy output of the central engine (0.18). $R_{\rm e}$ tracks the total and old components of the galaxy, which represent mass accumulated over billions of years. Since $L_{\rm AGN}$ correlates much more weakly with these total masses (0.45 for $M_*$) than with their young counterparts, it follows that it would also correlate weakly with $R_{\rm e}$. The data suggests that a galaxy's physical size is not a prerequisite for high AGN activity. A small, compact galaxy undergoing a significant gas inflow or starburst can host a very luminous AGN, while a massive, physically large galaxy may have a dormant central black hole if it lacks the young gas and stellar components required for fueling.

\item The perfect correlation (1.00) between two parameters are mathematically coupled or functionally dependent within the specific \texttt{CIGALE} model used to generate this data.

\end{itemize}

\subsection*{Final list of parameters correlating with $L_{\rm{AGN}}$}
We find that the instantaneous $({\rm SFR}_{10\,{\rm Myr}}$), SFH age,  stellar mass ($M_\star$), stellar age (Age$_\star$), and dust luminosity (L$_{\rm{dust}}$) are the most significant, each exhibiting a strong correlation with AGN luminosity. These results indicate that AGN luminosity tracks the galaxy's current star-formation activity most closely, consistent with a scenario in which black hole accretion and star formation are jointly regulated by the availability of cold gas rather than one process directly driving the other. Dust luminosity, which in our SED decomposition traces only the reprocessed radiation from stars, further supports this picture that their strong observed correlation cannot be attributed to a shared reprocessing origin, and instead reflects a genuine physical coupling between the  star-formation-heated dust reservoir and AGN accretion, both fueled by a common supply of cold gas.

\begin{figure*}[!ht]
    \centering
    \includegraphics[width=0.99\textwidth]{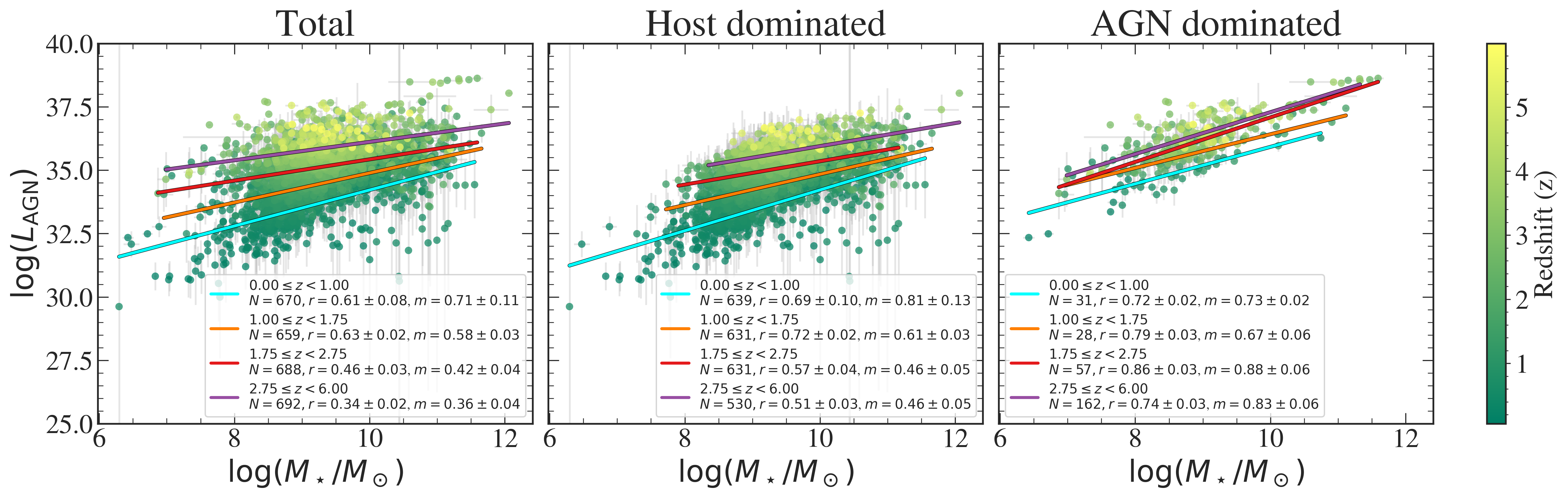}
  \includegraphics[width=0.99\textwidth]{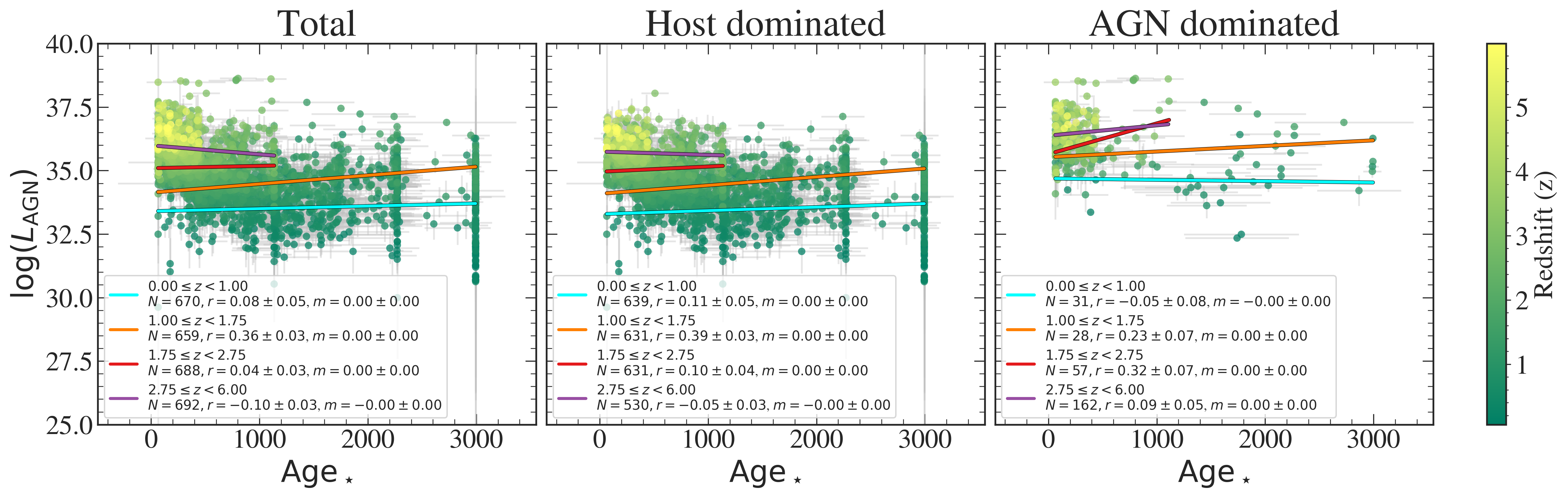}
    \caption{Same as Figure.~\ref{fig:vssfr} but AGN luminosity versus stellar mass ($M_\star$) and stellar age (${\rm Age}_\star$) [Myr].}
    \label{fig:vsstellar}
\end{figure*}
\begin{figure*}[!ht]
    \centering
    \includegraphics[width=0.99\textwidth]{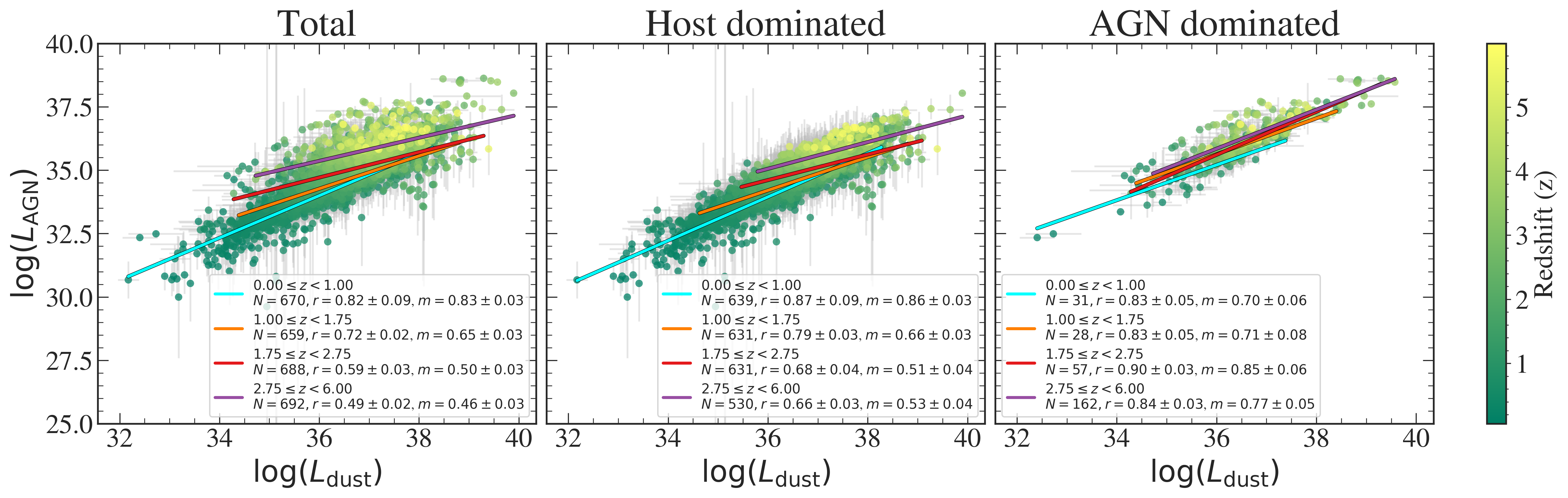}
    \caption{Same as Figure.~\ref{fig:vssfr} but AGN luminosity versus dust luminosity ($L_{\rm dust}$).}
    \label{fig:vsdust}
\end{figure*}

\section{Relationships Between $L_{\rm AGN}$ and Physical Galaxy Parameter}
\label{sec:evol}
For each parameter $X$, we construct a three-panel scatter plot of
$L_{\rm AGN}$ versus $X$, consisting of

\begin{itemize}
    \item the total sample,
    \item host-dominated galaxies ($f_{\rm AGN}\le0.2$), and
    \item AGN-dominated galaxies ($f_{\rm AGN}>0.2$).
\end{itemize}

The host/AGN classification is based on the \texttt{CIGALE} AGN fraction, adopting a threshold of $f_{\rm AGN}=0.2$. This threshold is chosen because it typically exceeds the measurement uncertainties and corresponds to the regime where AGN emission begins to contribute significantly to the observed spectral energy distribution. Lower thresholds are more susceptible to contamination from star formation and measurement noise, whereas higher thresholds (e.g. $f_{\rm AGN}>0.3$--$0.4$) would preferentially exclude composite systems and reduce the sample size.

To assess whether the observed correlations are driven primarily by
redshift-dependent selection effects, such as the Malmquist bias, we divide the sample into four approximately equal-number redshift bins constructed from the full sample. This allows the relationships between $L_{\rm AGN}$ and the host-galaxy properties to be examined over a narrower range of luminosity distance, thereby reducing the impact of the strong redshift dependence present in the full sample. Within each redshift bin, we perform an independent linear least-squares regression and compare the fitted relations across redshift. Although this approach does not completely remove selection effects, it provides a direct test of whether the observed trends persist within individual redshift intervals.

In each panel (Figure.~\ref{fig:vssfr}--\ref{fig:vsdust}), galaxies are color-coded by redshift, and the best-fitting linear relation for each redshift bin is over-plotted. The best-fit slopes ($m$) and Pearson correlation coefficient ($r$) and number of sources ($N$) for each redshift interval is also reported . Their uncertainties are estimated through Monte Carlo propagation of the reported CIGALE parameter uncertainties. The median of the perturbed distributions is not adopted as the fiducial value, since the CIGALE uncertainties represent posterior widths rather than independent measurement errors. 

\paragraph{Star Formation Rate (${\rm SFR}_{10 \mathrm{Myr}}$)}
The top panel of Figure.~\ref{fig:vssfr} shows the relation between $L_{\rm AGN}$ and ${\rm SFR}_{10 \mathrm{Myr}}$. A positive correlation is present in every redshift bin for the total, host-dominated, and AGN-dominated samples (Table.~\ref{tab:corr}), indicating that the connection between black hole accretion and recent star formation persists after dividing the sample into narrower redshift intervals. However, the correlation strength decreases with increasing redshift for the total sample, from $r=0.69\pm0.07$ at $z<1$ to $r=0.49\pm0.02$ at $2.75<z<6$, with the corresponding slope decreasing from $0.90\pm0.05$ to $0.60\pm0.04$. This evolution may reflect a combination of redshift-dependent selection effects, changes in the dynamic range of the observed population, and increasing stochasticity between recent SFR and instantaneous black-hole accretion. In contrast, the AGN-dominated population consistently exhibits much stronger correlations ($r\simeq0.80$--0.90) with slopes close to unity ($m\simeq0.86$--1.07), suggesting that AGN luminosity scales nearly linearly with the recent star formation rate once the AGN contributes significantly to the infrared emission. The host-dominated galaxies show intermediate correlations ($r\simeq0.64$--0.73), reflecting the larger contribution of stellar processes to the observed infrared luminosity. The persistence of the relation within individual redshift bins indicates that the observed $L_{\rm AGN}$--SFR connection is not solely a consequence of the overall increase in luminosity with redshift, although selection effects cannot be completely excluded.

\paragraph{SFH Age}
The bottom panel of Figure.~\ref{fig:vssfr} presents the relation between $L_{\rm AGN}$ and the SFH age. Unlike the recent star formation rate, no consistent trend is observed across redshift. The Pearson correlation coefficients are generally weak ($|r|\lesssim0.4$) and often statistically insignificant, particularly for the AGN-dominated sample. The sign of the correlation also changes with redshift, being weakly positive at intermediate redshifts and weakly negative in the highest-redshift bin. The fitted slopes are effectively zero in all cases, demonstrating that AGN luminosity is largely insensitive to the characteristic SFH age once redshift is controlled. These results suggest that the apparent anti-correlation observed in the full sample is primarily driven by the combined evolution of galaxy age and AGN luminosity across cosmic time rather than by a direct physical connection.

\paragraph{Stellar Mass ($M_\star$)}
The top panel of Figure.~\ref{fig:vsstellar} shows a positive correlation between AGN luminosity and stellar mass in every redshift interval. For the total sample, the correlation decreases from $r=0.61\pm0.08$ at low redshift to $r=0.34\pm0.02$ at the highest redshifts, accompanied by a reduction in slope from $0.71\pm0.11$ to $0.36\pm0.04$. In contrast, the AGN-dominated galaxies consistently exhibit stronger correlations ($r\simeq0.72$--0.86) and steeper slopes ($m\simeq0.67$--0.88), while the host-dominated population shows intermediate values ($r\simeq0.51$--0.72). This behavior is consistent with the well-established relationship between galaxy stellar mass and black hole mass \citep{Li_2021, Yang2025}. The stronger relation in AGN-dominated systems suggests that, when the AGN contributes substantially to the observed emission, the measured luminosity more directly traces black hole growth. The gradual weakening of the relation in the total sample with increasing redshift indicates that part of the global correlation arises from redshift-dependent sample evolution, although a significant intrinsic connection remains within individual redshift bins.

\paragraph{Stellar Age (${\rm Age}_\star$)}
The bottom panel of Figure.~\ref{fig:vsstellar} shows that stellar age exhibits little correlation with AGN luminosity once the sample is divided by redshift. Most Pearson coefficients satisfy $|r|<0.4$, with fitted slopes that are statistically consistent with zero. Similar to the SFH age, the sign of the correlation changes between redshift bins, and no monotonic behavior is observed for either the host- or AGN-dominated samples. These results indicate that the average stellar age of the galaxy is not a primary driver of AGN luminosity after accounting for cosmic evolution, implying that any trend present in the full sample is largely produced by the strong evolution of galaxy populations with redshift.

\paragraph{Dust Luminosity ($L_{\rm dust}$)}
A strong positive correlation between $L_{\rm AGN}$ and dust luminosity is 
observed in every redshift interval (Figure.~\ref{fig:vsdust}). The total 
sample shows a gradual decline in correlation strength from $r=0.82\pm0.09$ 
at $z<1$ to $r=0.49\pm0.03$ at $2.75<z<6$. Similar to the $L_{\rm AGN}$--${\rm SFR}_{10\,\mathrm{Myr}}$ relation, this decline may reflect a combination of redshift-dependent selection effects, changes in the dynamic range of the observed population, and increasing stochasticity between star-formation-heated dust emission and instantaneous AGN accretion at higher redshift, since $L_{\mathrm{dust}}$ and $L_{\mathrm{AGN}}$ are derived from independent, non-overlapping star-formation and AGN components in our SED decomposition. However, the AGN-dominated galaxies consistently maintain strong correlations ($r\simeq0.83$--0.90). The fitted slopes are systematically steeper for the AGN-dominated population ($m\simeq0.70$--0.85) than for the host-dominated sample ($m\simeq0.51$--0.86), indicating that in AGN-dominated systems, star-formation-heated dust emission scales more tightly with AGN power, consistent with both processes being fueled by a shared, actively growing gas reservoir when the AGN dominates the infrared energy budget. In host-dominated galaxies, the star-formation-heated dust reservoir is set largely by the galaxy's independent star-formation history, introducing additional scatter relative to AGN accretion and resulting in weaker correlations. Nevertheless, the persistence of a significant $L_{\rm AGN}$--$L_{\rm dust}$ relation within each redshift interval -- despite the two quantities coming from non-overlapping energy budgets -- suggests that the coupling between AGN activity and star-formation-driven dust heating is intrinsic and not solely produced by the common increase of luminosity with redshift.

Overall, correlations strengthen when the measured quantity directly traces AGN energy output (e.g., SFR, dust luminosity), and weaken when the parameter reflects integrated or indirect properties (e.g., SFH age). Overall, these trends support a scenario in which AGN activity is closely linked to gas availability, star formation, and host galaxy mass.
Another noticeable point is the host-dominated sample represents the general galaxy population, which includes everything from young to old galaxies that have ceased AGN growth. The AGN-dominated systems are ``growth specialists''-they are almost exclusively young, gas-rich, and actively evolving systems, which naturally creates a cleaner, less scattered statistical relationship.

\section{Principal Component Analysis (PCA)}
\label{sec:pca}

We perform PCA on the selected parameters to identify the dominant axes of variation in the galaxy + AGN parameter space. Data are standardized using 
\begin{equation}
X' = \frac{X - \langle X \rangle}{\sigma_X},
\end{equation}
and the first two principal components (PC1, PC2) are extracted:
\begin{equation}
\mathbf{PC} = \mathbf{X'} \cdot \mathbf{V},
\end{equation}
where $\mathbf{V}$ contains the eigenvectors of the covariance matrix.

\begin{figure}[ht]
    \centering
    \includegraphics[width=0.45\textwidth]{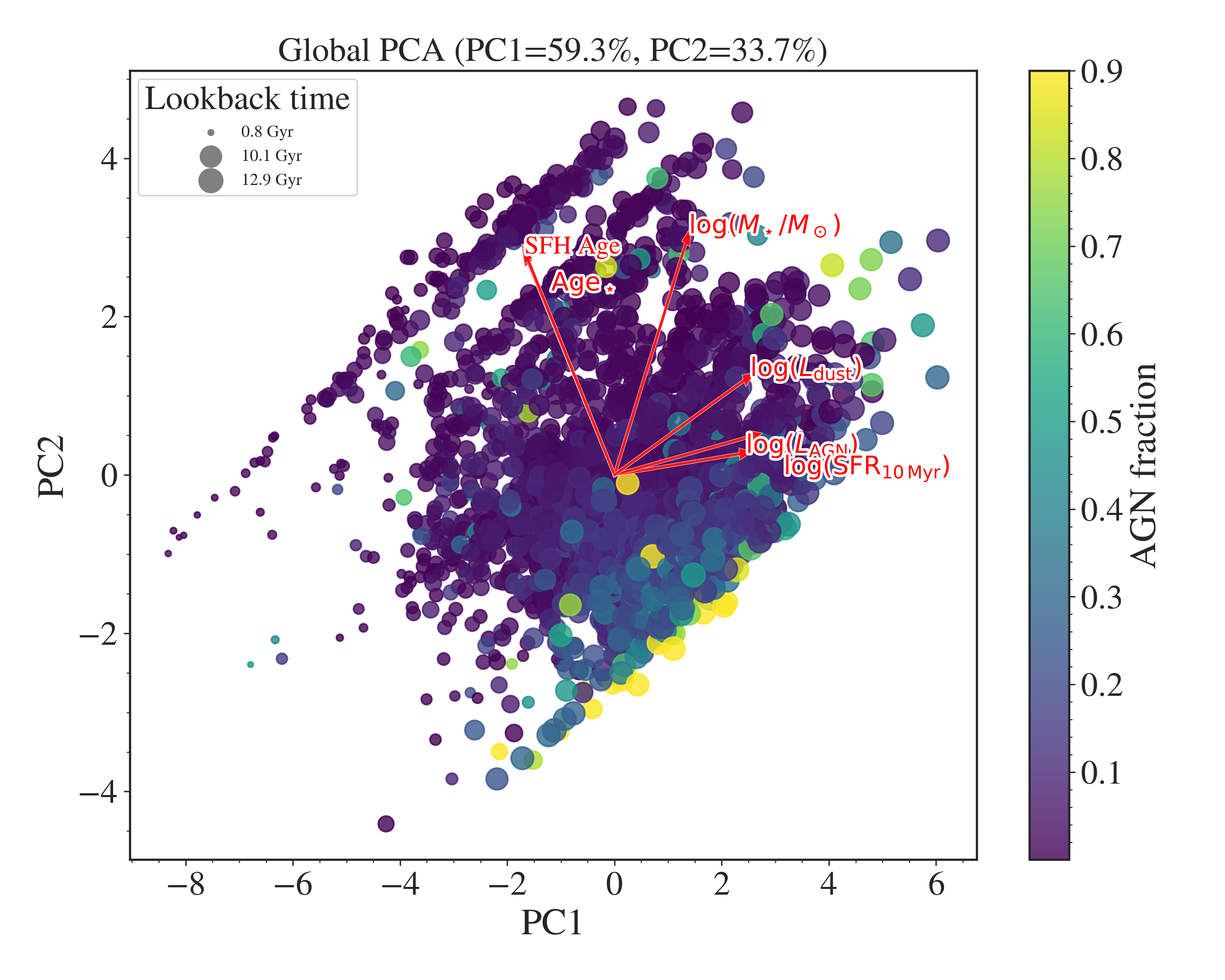}
       \caption{Global principal component analysis (PCA) of the full galaxy sample ($N=2709$), explaining 93\% of the total variance. PC1 (59.3\%) traces an activity axis defined by $L_{\rm AGN}$, ${\rm SFR}_{10\,{\rm Myr}}$, and $L_{\rm dust}$, while PC2 (33.7\%) represents an evolutionary axis where ${\rm Age}_*$ and SFH age are co-aligned and anti-correlated with activity. High AGN fractions populate the high-activity, young regime, with high-redshift systems driving the variance along PC1. To plot the arrows we plot the loadings multiplied by a factor of five for visual clarity.}
    \label{fig:allfrac}
\end{figure}

\begin{table*}
\centering
\begin{tabular}{lcc}
\hline
Parameter & Direction & Physical Interpretation \\
\hline
$L_{\rm AGN}$ & $+\,$PC1 & Traces instantaneous AGN activity \\
${\rm SFR}_{10\,\mathrm{Myr}}$ & $+\,$PC1 & Co-evolves with AGN fueling \\
${\rm SFH Age}$ & $-\,$PC1, $+\,$PC2 & Older systems are less active but more evolved \\
${\rm Age}_\star$ & $-\,$PC1, $+\,$PC2 & Older systems are less active but more evolved \\
$M_\star$ & $+\,$PC2 & Tracks cumulative galaxy growth \\
$L_{\rm dust}$ & $+\,$PC1, $+\,$PC2 & Links activity with dust content \\
\hline
\end{tabular}
\caption{Summary of PCA Vectors for the entire sample.}
\label{tab:global}
\end{table*}

\begin{figure*}[ht]
    \centering
    \includegraphics[width=0.49\textwidth]{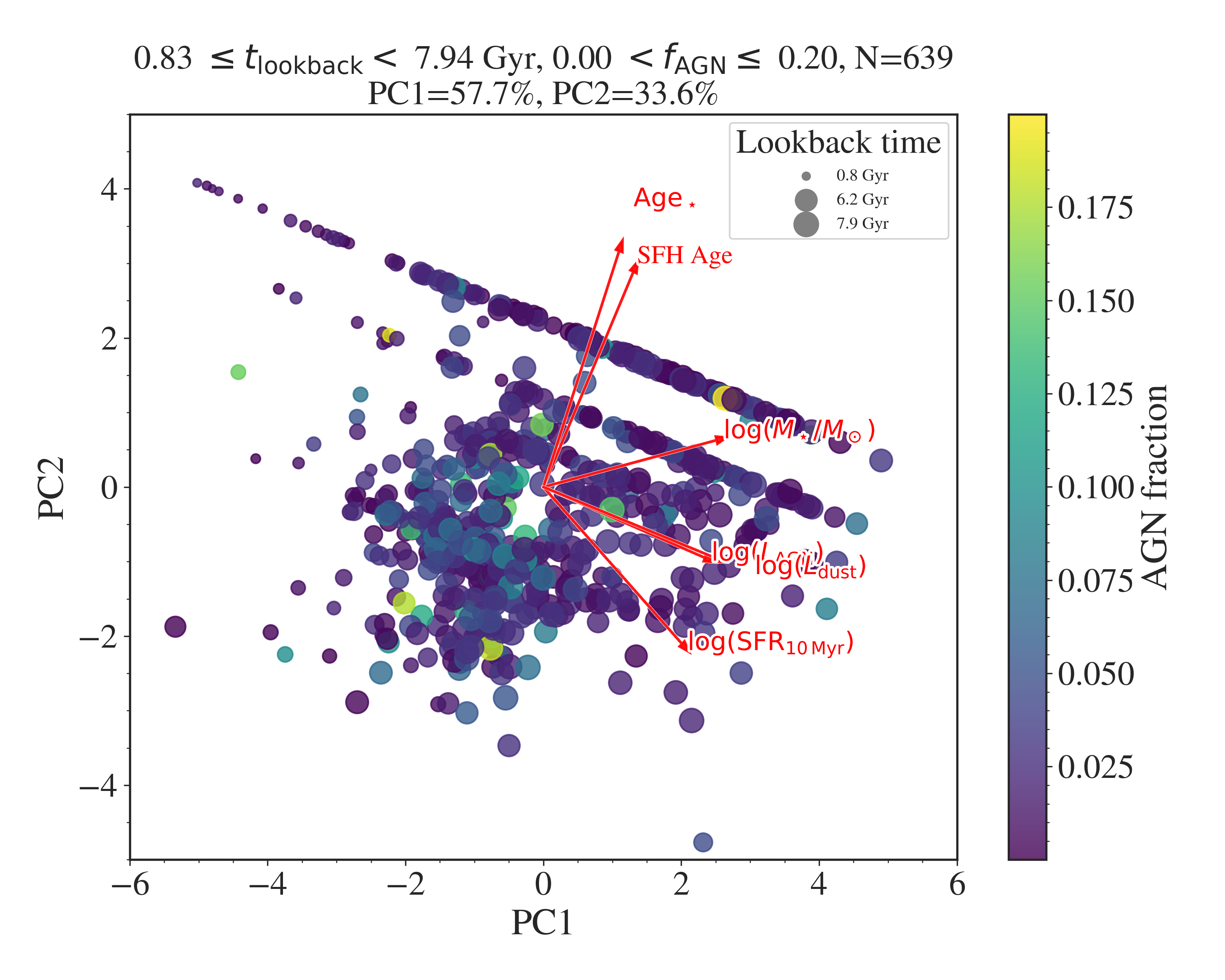}
       \includegraphics[width=0.49\textwidth]{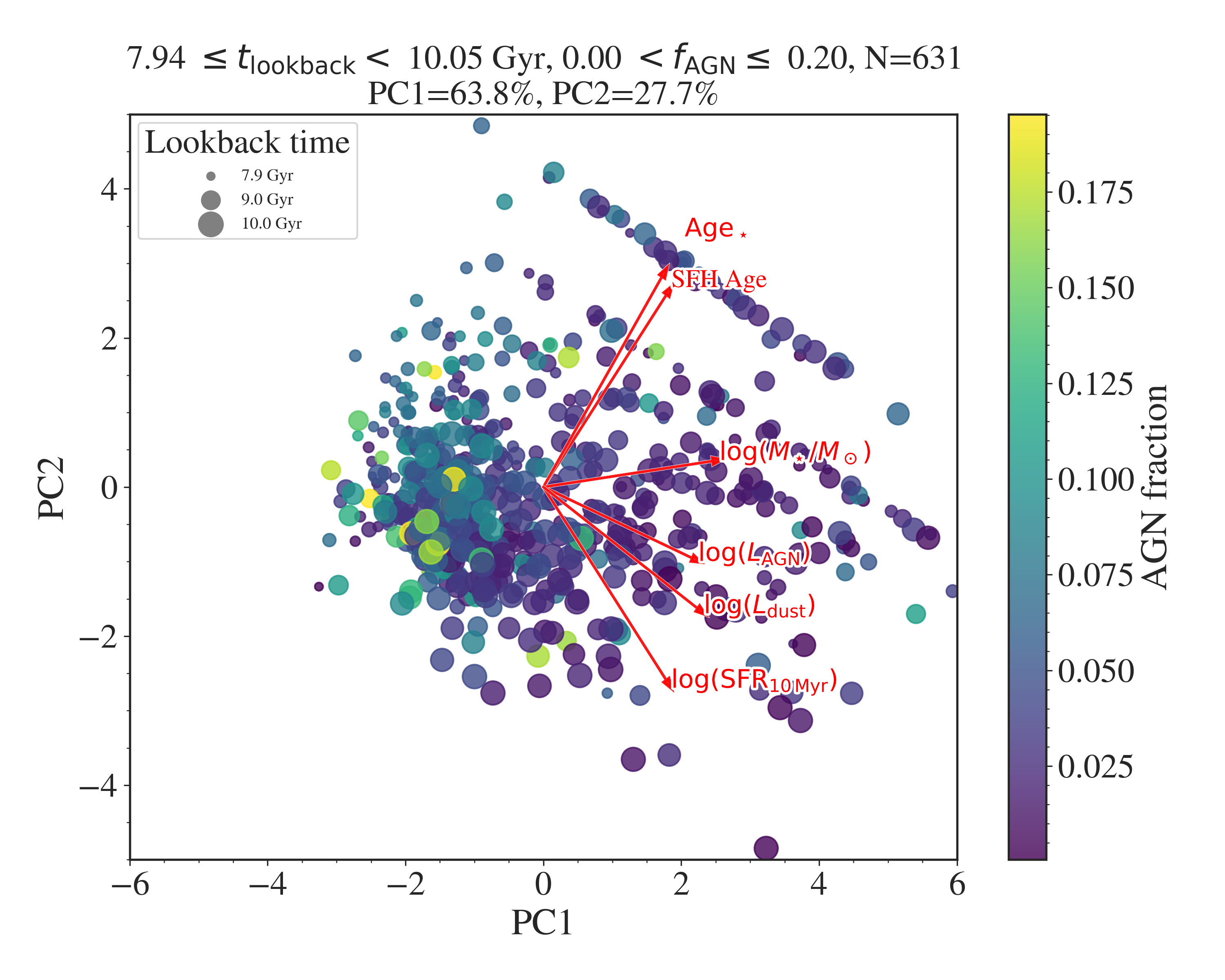}    \includegraphics[width=0.49\textwidth]{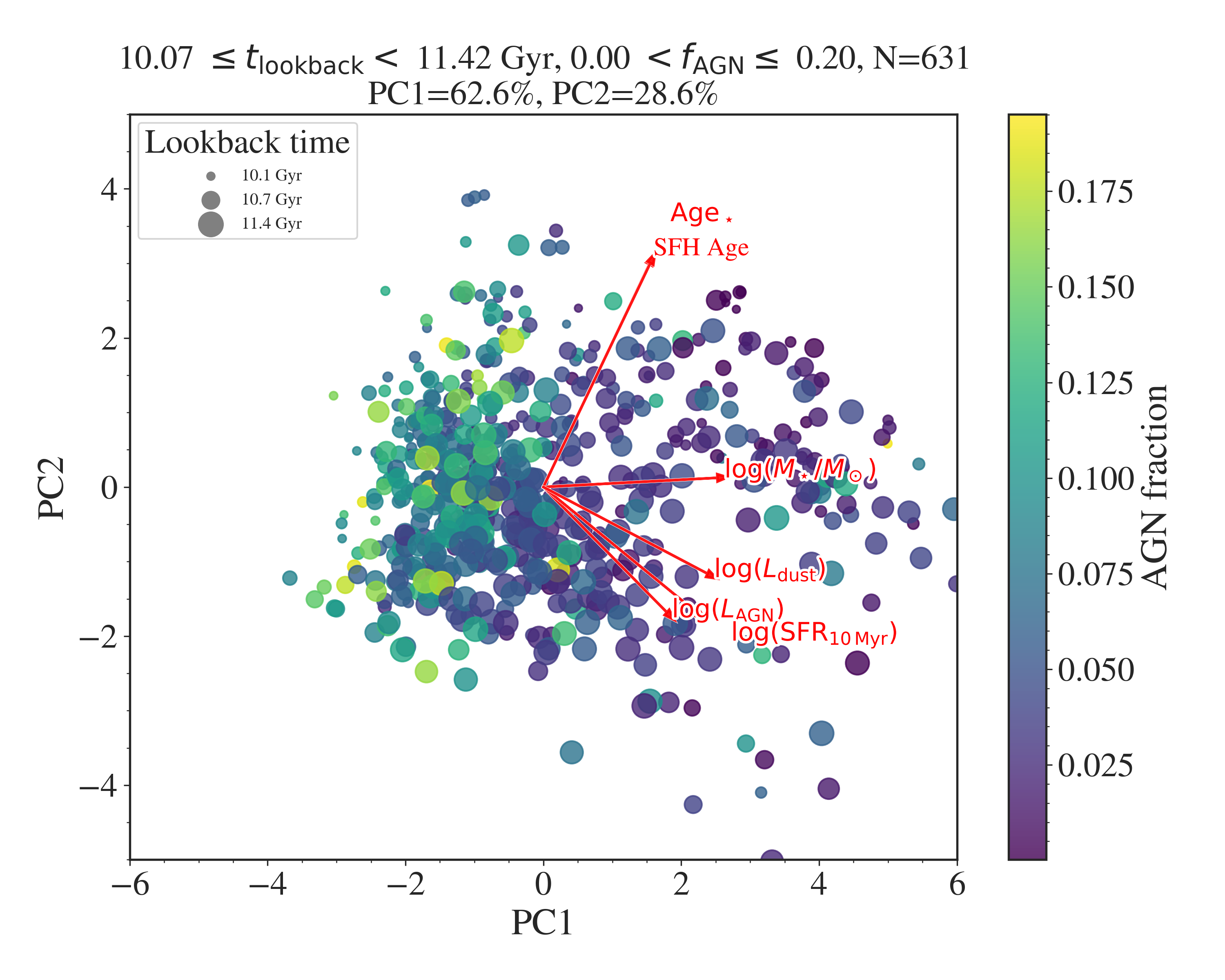}    \includegraphics[width=0.49\textwidth]{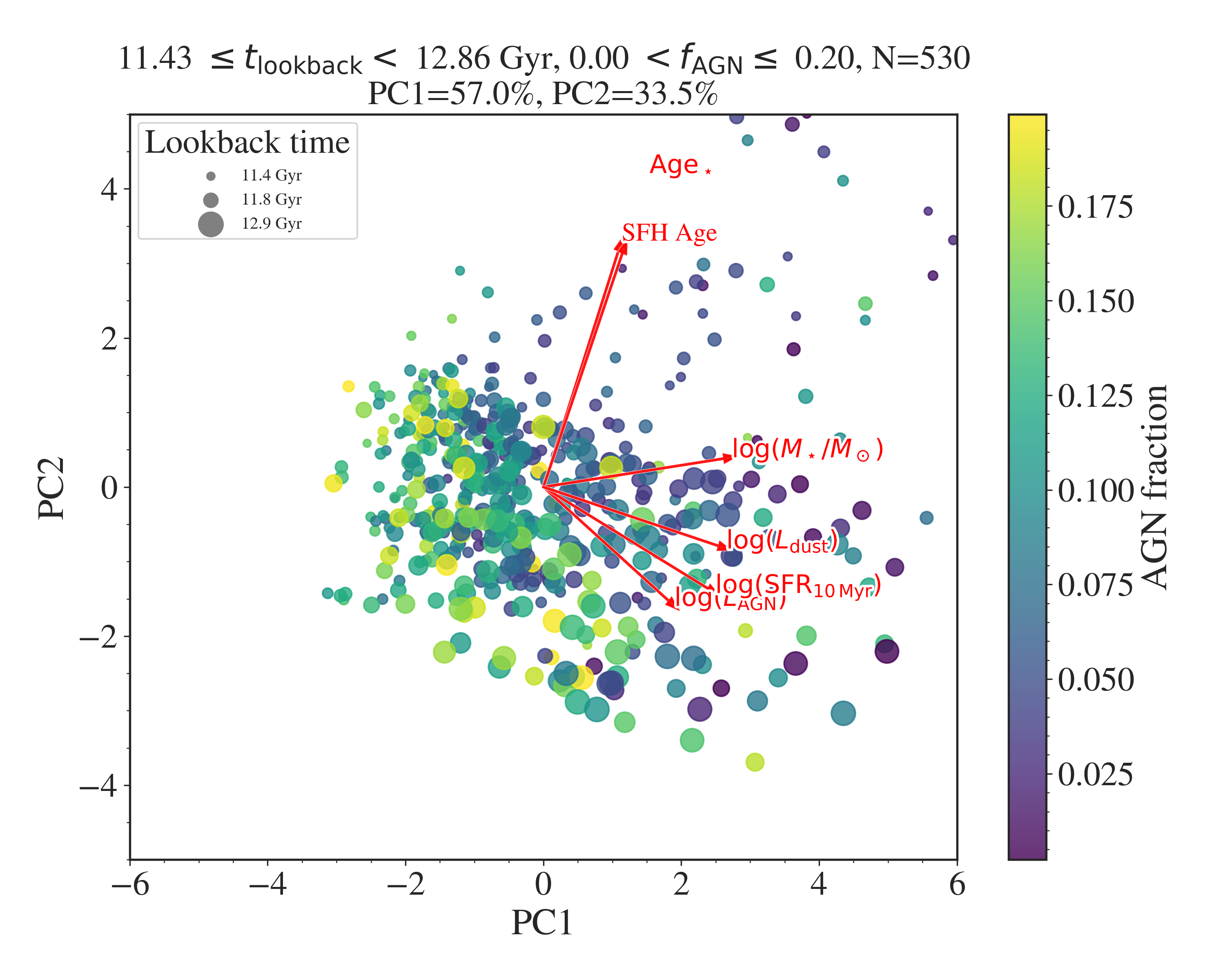}
       \caption{Principal Component Analysis (PCA) of physical properties for galaxies with low AGN fractions ($f_{\mathrm{AGN}} \leq 0.20$) across four cosmic epochs, ranging from the early Universe to the present day. PC1 represents the overall scale and activity of galaxies, driven primarily by stellar mass, star-formation rate (SFR), dust luminosity, and AGN luminosity. PC2 is primarily defined by the age of the stellar population. To plot the arrows, we plot the loadings multiplied by a factor of 5 for visual clarity.}
    \label{fig:frac0}
\end{figure*}

\subsection{Global Principal Component Analysis}
\label{sec:global}
The global PCA, shown in Figure.~\ref{fig:allfrac}, captures the joint behavior of the five primary galactic parameters and AGN luminosity, explaining $\sim93\%$ of the total variance.

The first principal component (PC1) is dominated by the current level of galactic activity. The vectors for $L_{\rm AGN}$ and ${\rm SFR}_{10\,\mathrm{Myr}}$ are strongly aligned along positive PC1, indicating that the primary driver of variance is the coeval fueling of black hole accretion and star formation. 
Dust luminosity also contributes positively, suggesting that more massive and evolved galaxies retain significant dust emission.

The second principal component (PC2) traces the integrated growth and evolutionary state of galaxies. Stellar mass dominates this axis, pointing toward positive PC2 and separating massive, evolved systems from less massive ones. SFH age and ${\rm Age_\star}$ completely overlaps and points toward negative PC1 and positive PC2, reflecting the decline of activity as galaxies evolve.

High AGN fractions (shown by color-bar) cluster at high PC1 and low PC2, indicating that AGN-dominated systems are highly active but not necessarily the most massive. Moreover, it again confirms that the high AGN fraction is a characteristic of younger systems that have not yet reached the ``mature'' or ``passive'' state occupied by the older galaxies at the top of the plot. Point size (lookback time) shows that high-redshift systems preferentially occupy the high-PC1 region. This trend may partly reflect a combination of intrinsic cosmic evolution and selection effects, since lower-luminosity (low-SFR or low-AGN) systems become increasingly difficult to detect at higher redshift in flux-limited survey. Nevertheless, within the detected population, early systems are predominantly found in the high-activity regime defined by PC1, but the most massive galaxies are typically found in the later universe after billions of years of growth, even if their central AGNs have become less luminous. A brief summary of the PCA is mentioned in Table~\ref{tab:global}.

To capture potential evolution with lookback time and AGN fraction, we bin galaxies by AGN fraction $f_{\rm AGN} > 0.2$ and quartiles of lookback time $t_{\rm lookback}$ corresponding to the redshift bins in previous section. PCA is performed independently in each lookback-time and AGN-fraction bin, and the resulting loadings are used to study the evolution of physical parameter correlations. The PCA results are visualised in the PC1--PC2 plane, where loadings are represented as vectors, point size scales with lookback time, and color encodes $f_{\mathrm{AGN}}$. The loading details are shown in Table~\ref {tab:pca_low_agn_evolution}.

\subsection{PCA Evolution of Host-Dominated Galaxies}
Similar to previous analysis (Section~\ref{sec:global}), we now analyze the relationships between key physical properties and AGN luminosity for host-dominated galaxies ($f_{\rm AGN} \leq 0.20$) and the evolution of relation from early cosmic times (high lookback time) to the present-day Universe. As can be seen from Figure.~\ref{fig:frac0}, across all epochs, despite the evolution in galaxy distribution, the orientations of the loading vectors remain broadly stable, with only modest variations in their relative magnitudes and angles. The vectors corresponding to $L_{\mathrm{AGN}}$, $L_{\mathrm{dust}}$, and $\mathrm{SFR}_{10\,\mathrm{Myr}}$ remain closely aligned across all epochs. This persistent alignment indicates a strong coupling between AGN activity, dust emission, and recent star formation. Across all epochs, galaxies with higher AGN fractions (highlighted in yellow) are consistently located on the opposite side of the PCA space relative to the stellar mass vector, oriented toward the positive PC1 axis. In massive galaxies (located on the far right of the plots), the host's own stellar and the dust emission is so high that even a luminous AGN might only account for a small percentage of the total IR output. Conversely, in less massive galaxies, it is ``easier'' for the AGN to dominate the total energy budget, resulting in a higher AGN fraction. Moreover, the stellar age vector consistently lies nearly orthogonal to the dust luminosity vector, indicating that stellar age represents a largely independent dimension from $L_{\mathrm{dust}}$ throughout cosmic time confirming the ``age-blind'' scenario.

We emphasize that differences between redshift bins reflect the evolution of the observed galaxy population under a flux-limited selection function, rather than a fully mass-complete census. Therefore, changes in PCA structure with redshift should be interpreted with caution, as they may reflect both intrinsic evolution and evolving selection effects.

At higher redshifts/epochs ($10.4$--$12.9$ Gyr), PC1 explains $57\%$ of the total variance, while the second principal component accounts for $33.5\%$. The galaxy distribution appears relatively clustered, with activity-related vectors, including SFR, AGN luminosity, and dust luminosity, oriented toward the lower-right quadrant (positive PC1, negative PC2). A noticeable point at this epoch is that the vector corresponding to $L_{\mathrm{AGN}}$ is very closely aligned with that of $\mathrm{SFR}_{10\,\mathrm{Myr}}$. This alignment suggests that, in the early Universe, black hole growth and star formation were tightly coupled processes, likely fueled by a common reservoir of abundant cold gas. A similar trend is also observed at slightly lower epoch ($10.07$--$11.4$ Gyr), with a bit more variance from PC1 of  approximately $\sim63\%$.

At intermediate lookback time ($7.9$--$10.1$ Gyr), PC1 accounts for approximately $\sim63\%$ of the variance, while PC2 contributes $\sim28\%$. In this bin, the galaxy distribution exhibits a broader spread in loading-vector orientations compared to the highest-redshift sample. This indicates that during this period, global scaling properties such as stellar mass also emerge as the primary drivers of variance within the galaxy population. However, there is a visible fanning out or widening of the angles between the loading vectors. In the earliest epochs, properties like mass, star formation, and luminosity were tightly grouped, meaning they evolved in lockstep. The ``fanning out'' at intermediate times signifies that these physical properties began to decouple, as galaxies followed more diverse evolutionary paths where, for example, a high stellar mass did not always guarantee a high star formation rate. 

In the lowest-lookback-time bin or towards present Universe, PC1 explains $57.7\%$, of the variance, while PC2 contributes $33.6\%$. The PCA distribution in this regime exhibits a more pronounced diagonal structure, where the larger dynamic range of detected galaxies allows clearer separation between star-forming (lower-right region) and quiescent populations  (upper-left region).  This diagonal structure suggests a population of quiescent, evolved galaxies characterized by older stellar ages (high PC2) and relatively low activity. Along the ridge, galaxies transition toward higher stellar mass (increasing PC1) and dominated by old stellar populations. The galaxies populating this sequence exhibit very low AGN fractions ($f_{\mathrm{AGN}} \approx 0$), indicating minimal ongoing black hole accretion or star formation. This ridge is also visible at 7.9-10.1 Gyr but less densely populated and more dispersed. During this time, more galaxies were still actively forming stars, so the population of truly ``dead'' or quiescent massive galaxies was still in the process of forming. In this current epoch, the $L_{\mathrm{AGN}}$ vector shifts to align almost entirely with $L_{\mathrm{dust}}$. This transition indicates that, in the more recent Universe, AGN activity is more strongly associated with dust-obscured environments rather than with immediate, intense star formation. This alignment may partly reflect the preference of the adopted CIGALE parameter space toward obscured (Type$\sim$2-like) AGN solutions, which can couple the inferred AGN luminosity more strongly to the IR dust emission. 
We also caution that the sharpness of this diagonal sequence could be because of CIGALE physical parameters over a finite model grid (e.g., discrete steps in $\tau$, stellar age, and AGN fraction), galaxies with lower signal-to-noise photometry can be preferentially assigned to a limited set of grid values rather than a continuum, which, under a linear projection such as PCA, can sharpen an otherwise more diffuse trend into a tighter apparent track. We checked that the ridge is populated by a broad range of stellar mass and dust luminosity values rather than being concentrated at grid extrema, and that its defining feature, uniformly low  $f_{\mathrm{AGN}}$, is not itself a parameter subject to grid discretizations in the same way, supporting a physical origin for the ridge.  Moreover, stellar age and SFH age become more dispersed at these epochs. These trends reflects the increasing diversity of the current Universe.

In the early epochs, the observed high-redshift population appears more clustered in PCA space, consistent with a selection toward luminous, actively star-forming systems. By the lower epochs, the population had split into the bimodal distribution seen in the present universe: some galaxies continued forming stars and remained ``young'', while others ceased star formation billions of years ago and became ``old''.

\begin{figure*}[ht]
    \centering
    \includegraphics[width=0.49\textwidth]{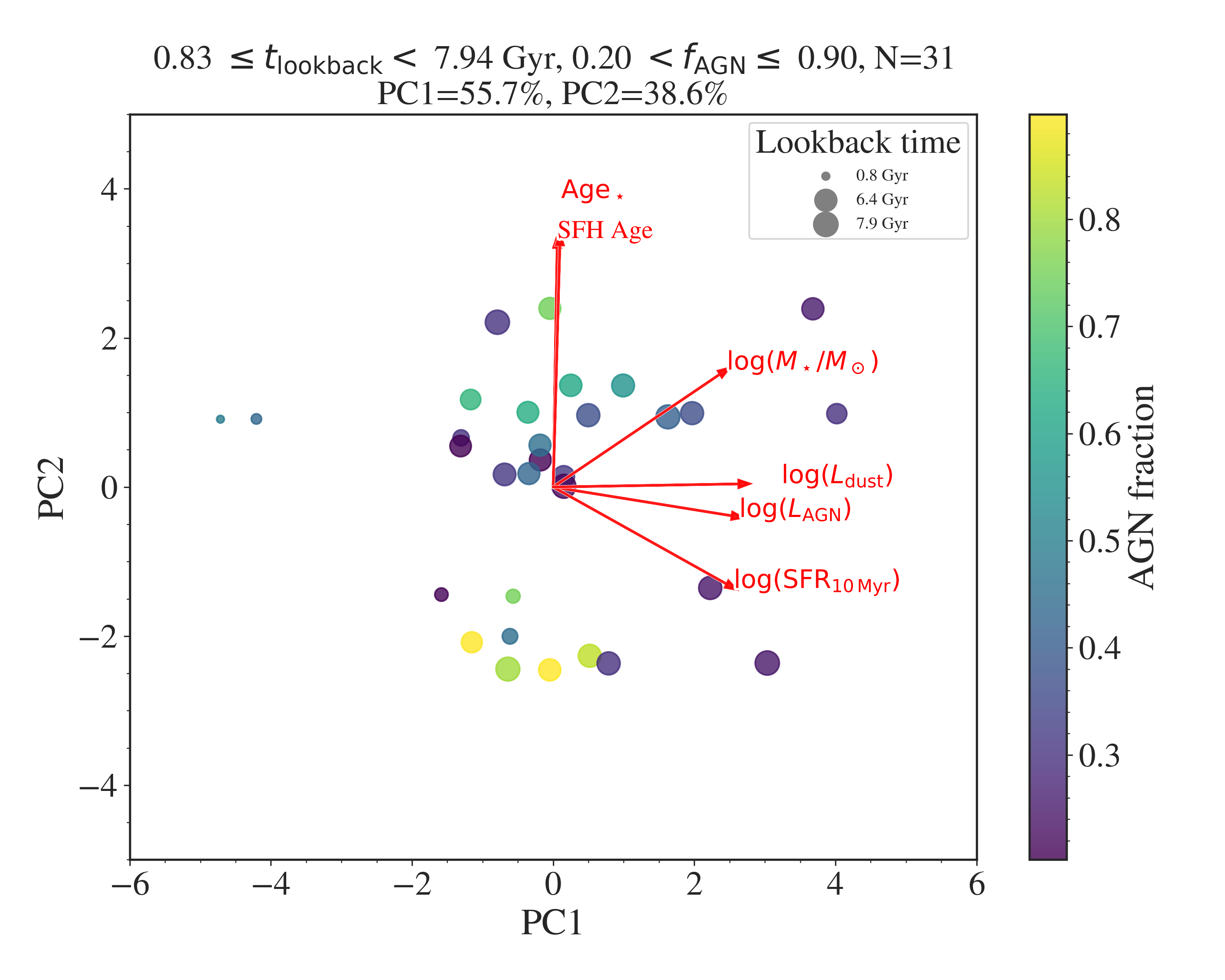}
       \includegraphics[width=0.49\textwidth]{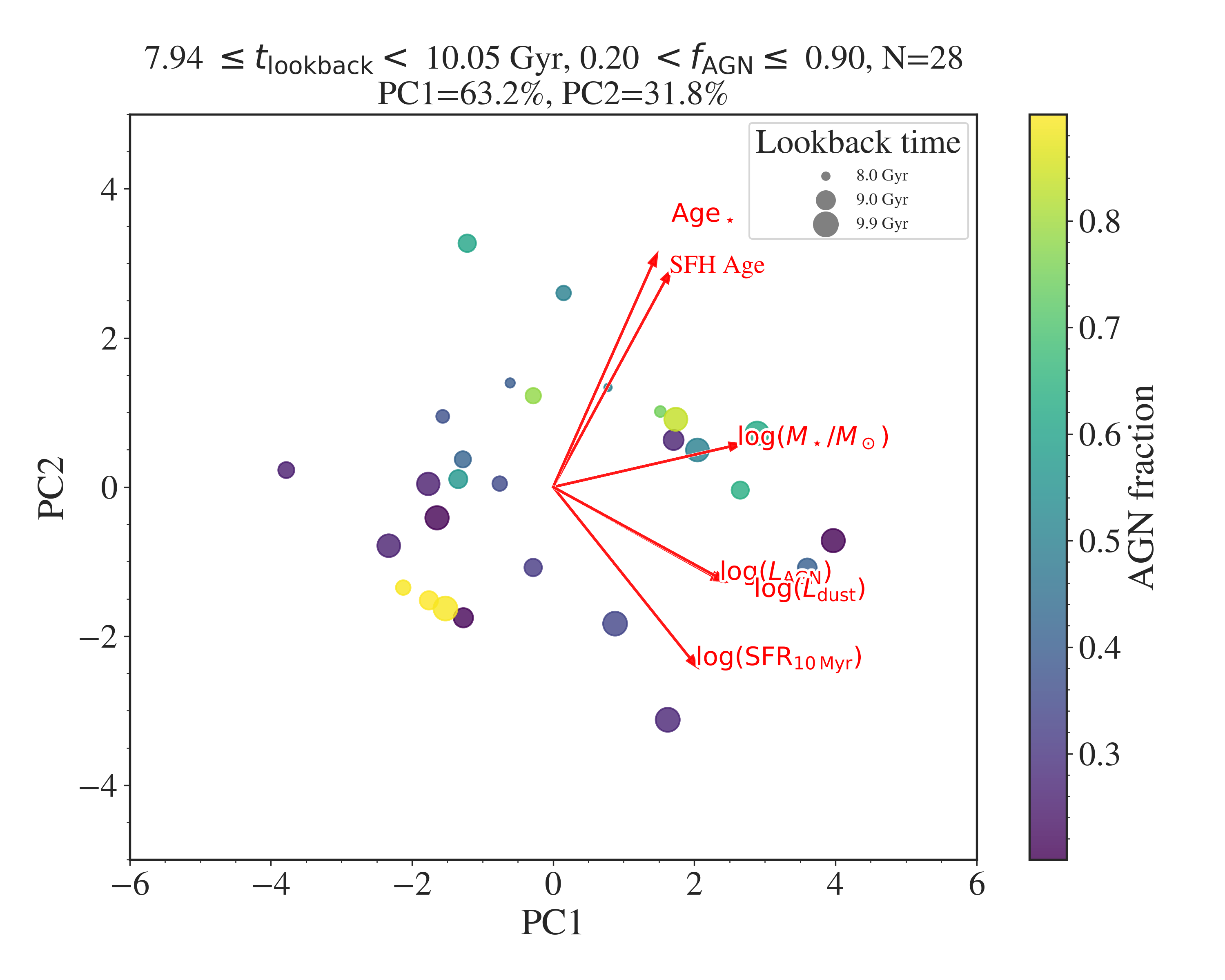}    \includegraphics[width=0.49\textwidth]{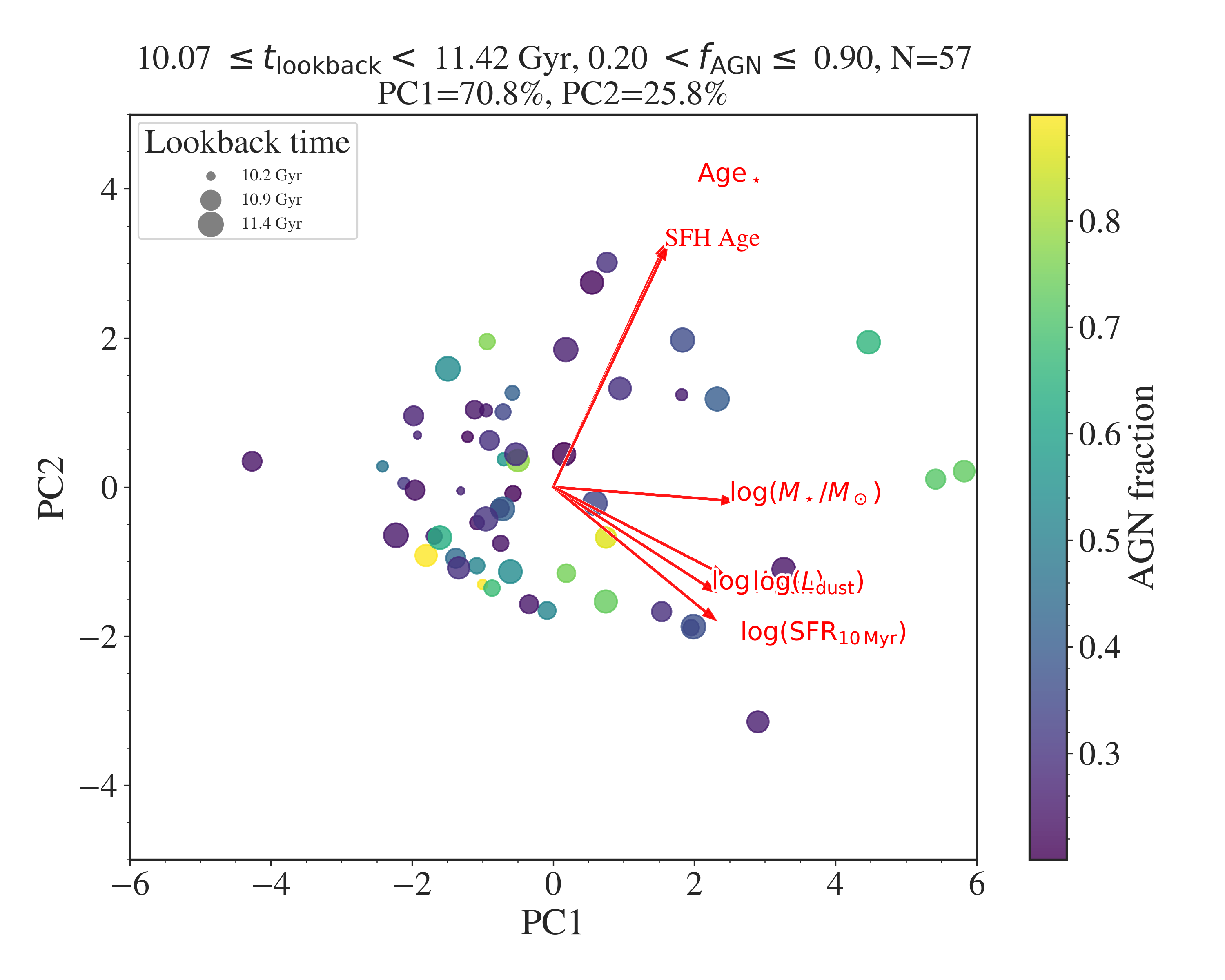}    \includegraphics[width=0.49\textwidth]{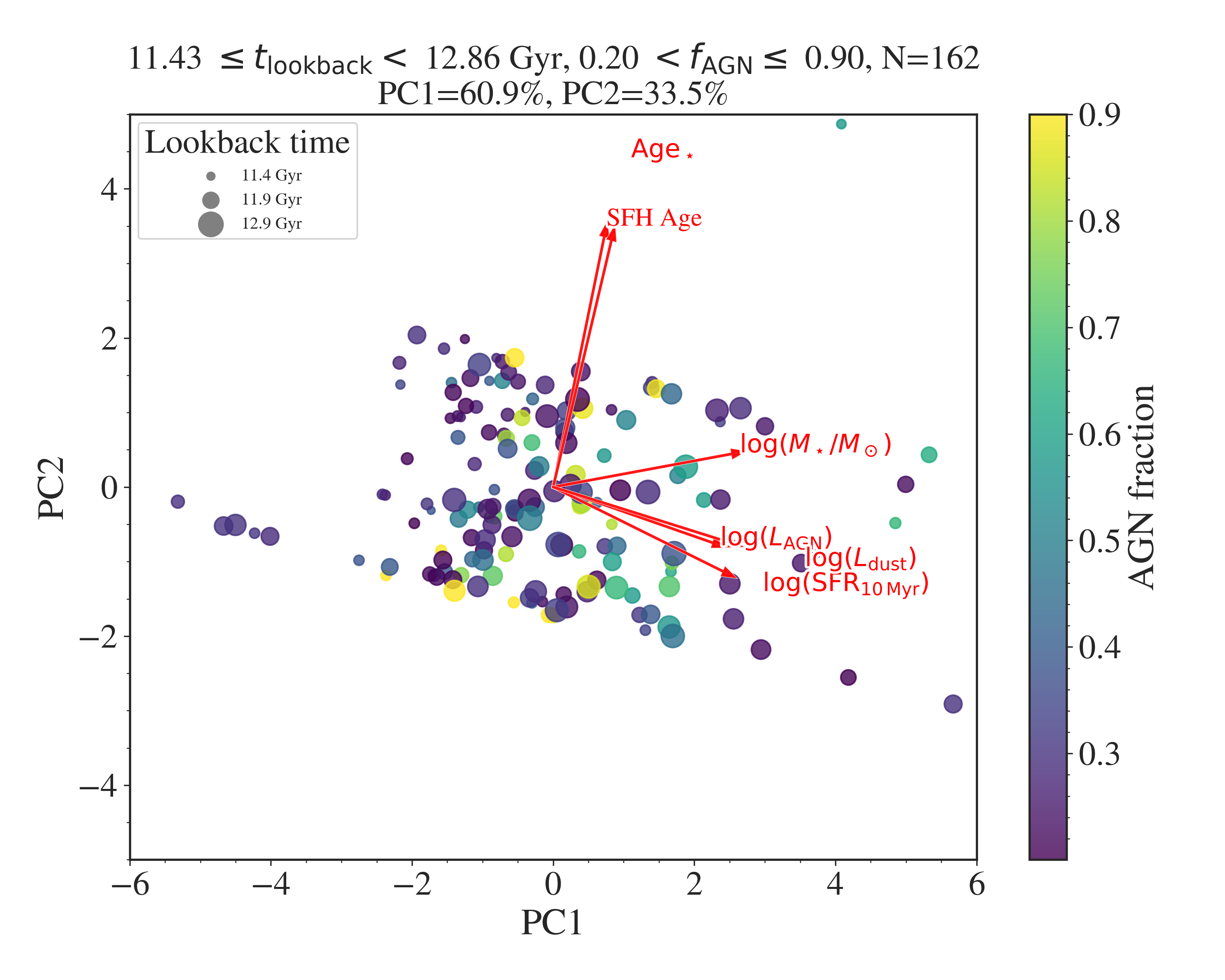}
       \caption{Same as Figure.~\ref{fig:frac0} but for AGN-dominated sample ($f_{\rm{AGN}} > 0.2$).}
    \label{fig:frac1}
\end{figure*}

\begin{table*}[!ht]
\centering
\caption{PCA loadings and their cosmic evolution.}
\begin{tabular}{lrrrrrrrr}
\hline
& \multicolumn{4}{c}{PC1} & \multicolumn{4}{c}{PC2}\\
Lookback time (Gyr) & 0.80-7.94 & 7.9-10.06 & 10.06-11.43 & 11.43-12.86 & 0.80-7.94 & 7.9-10.06 & 10.06-11.43 & 11.43-12.86 \\
\hline
& \multicolumn{8}{c}{Host-dominated}\\
\hline
$\log(L_{\rm AGN})$ & 0.469 & 0.432 & 0.358 & 0.365 & -0.190 & -0.191 & -0.336 & -0.307\\
$\log({\rm SFR}_{10\,{\rm Myr}})$ & 0.402 & 0.356 & 0.447 & 0.479 & -0.421 & -0.518 & -0.338 & -0.273\\
SFH Age & 0.261 & 0.357 & 0.307 & 0.218 & 0.578 & 0.516 & 0.598 & 0.636\\
$\log(M_\star/M_\odot)$ & 0.503 & 0.491 & 0.505 & 0.525 & 0.126 & 0.070 & 0.025 & 0.077\\
${\rm Age}_\star$ & 0.220 & 0.344 & 0.310 & 0.230 & 0.633 & 0.564 & 0.600 & 0.629\\
$\log(L_{\rm dust})$ & 0.499 & 0.447 & 0.476 & 0.510 & -0.190 & -0.326 & -0.232 & -0.159\\
\hline
& \multicolumn{8}{c}{AGN-dominated}\\
\hline
$\log(L_{\rm AGN})$ & 0.507 & 0.453 & 0.432 & 0.455 & -0.077 & -0.239 & -0.264 & -0.150\\
$\log({\rm SFR}_{10\,{\rm Myr}})$ & 0.492 & 0.388 & 0.434 & 0.484 & -0.260 & -0.460 & -0.336 & -0.228\\
SFH Age & 0.011 & 0.317 & 0.304 & 0.145 & 0.642 & 0.553 & 0.622 & 0.674\\
$\log(M_\star/M_\odot)$ & 0.473 & 0.501 & 0.480 & 0.508 & 0.303 & 0.109 & -0.035 & 0.089\\
${\rm Age}_\star$ & 0.019 & 0.281 & 0.308 & 0.166 & 0.649 & 0.597 & 0.616 & 0.664\\
$\log(L_{\rm dust})$ & 0.526 & 0.462 & 0.456 & 0.501 & 0.008 & -0.239 & -0.223 & -0.149\\
\hline
\end{tabular}
\label{tab:pca_low_agn_evolution}
\end{table*}

\subsection{PCA Evolution of AGN-Dominated Systems}
Similar to the previous section, we now analyze the evolution of physical drivers across cosmic time for the high AGN fraction population ($0.20 < f_{\mathrm{AGN}} < 0.90$) as shown in Fig~\ref{fig:frac1}. Similar to host-dominated systems, the general trend is consistent, PC1 is consistently driven by the overall \textit{scale and activity} of galaxies, including stellar mass, SFR, dust luminosity, and AGN luminosity. The second principal component is dominated by stellar age and SFH age. However, in AGN-dominated systems, galaxies with the highest AGN fractions do not necessarily tend to cluster toward the negative PC1 axis, opposite to the stellar mass vector. This suggests that the relative AGN IR contribution in AGN-dominated systems doesn't depend on stellar mass within this selected population. Interestingly, the vectors for stellar age and SFH age are nearly identical in this sample. The complete overlap indicates that strong AGN activity is a marker for a specific, consistent evolutionary state where star-formation history and current stellar age remain consistently aligned within the observed sample.

In the earliest epoch, PC1 explains $60.9\%$ of the total variance. Galaxies form a dense, centralized distribution with a slight diagonal structure at lower ages. This diagonal trend, within the detected population at high redshift, shows that for these ``younger'' galaxies (those with the most recent star formation), there is a strong correlation between stellar mass and current activity levels. In this epoch, as a galaxy's mass increased, its star formation and AGN activity typically increased along with it, creating that specific diagonal tilt.

In the intermediate epoch, PC1 accounts for $\sim 63-70\%$ of the variance, while PC2 contributes $\sim25-31\%$.  The galaxy distribution becomes more elongated along PC1, reflecting increasing differentiation in galaxy properties. The ``fanning out'' phase, similar to host-dominated systems, the vectors for stellar age and SFH age are slightly separated, and the angular separation between stellar mass and SFR increases, signature of demographic diversity.

In the lowest-lookback-time bin,  PC1 explains $55.7\%$ of the variance, while PC2 contributes $38.6\%$. The activity-related vectors exhibit a broader angular separation. $L_{\mathrm{AGN}}$ and $L_{\mathrm{dust}}$ remain closely aligned, while $\mathrm{SFR}_{10\,\mathrm{Myr}}$ shifts toward more negative PC2 values and becomes nearly orthogonal to stellar mass, indicating a weaker correlation within the observed sample. As mentioned before, given that the adopted \texttt{CIGALE} parameter space preferentially selects obscured (Type~2-like) AGN solutions, the close alignment between $L_{\mathrm{AGN}}$ and $L_{\mathrm{dust}}$ may partly reflect the stronger coupling between inferred AGN luminosity and IR dust emission within the SED-fitting framework. 

To understand whether the conclusions are robust we perform PCA across three distinct volume-limited subsamples (Appendix: \ref{sec:appendix}) and found consistent results.

\section{Summary and Conclusions}
\label{sec:conc}

Based on the correlation analysis of a multi-wavelength catalog of 2,709 sources, we identify galaxy parameters that are most strongly connected to SMBH growth. We remove redundant information while preserving key correlations and explore the joint parameter space via PCA, both globally and in bins of lookback time and AGN fraction.

The statistical analysis demonstrates that $L_{\mathrm{AGN}}$ is a strong indicator of a young and active galaxy state:

\begin{itemize}
    \item Strong positive correlations with dust luminosity, stellar mass and SFR, even in various redshift bins ($r \approx 0.49$--$0.90$). .
    \item Statistically insignificant correlation with stellar age when redshift-dependence is taken into account. 
\end{itemize}

These trends indicate that luminous AGN are most prevalent in the early, gas-rich Universe and are typically associated with galaxies undergoing intense star formation and hosting younger stellar populations.

From PCA analysis, we find that across all epochs and AGN fractions, galaxy properties are governed by two primary dimensions (i)
\textbf{PC1 (scale and activity):} Dominated by  star-formation rate (SFR), dust luminosity, stellar mass, and AGN luminosity, (ii) \textbf{PC2 (stellar age):} Primarily defined by the age of the stellar population.
Although AGN luminosity ($L_{\mathrm{AGN}}$) correlates positively with stellar mass ($r \approx 0.3$--$0.7$), the fractional AGN contribution ($f_{\mathrm{AGN}}$) shows a contrasting trend. High-$f_{\mathrm{AGN}}$ systems are preferentially located opposite to the stellar mass vector in PCA space. This indicates that the relative energetic dominance of AGN activity is greatest in lower-mass galaxies, where the AGN can more effectively outshine the host galaxy emission.

The distribution of galaxy populations in the PCA space reveals distinct evolutionary pathways. 
\textbf{Host-dominated systems:} These galaxies exhibit a pronounced structural evolution, transitioning from a centralized distribution at early epochs to a clear bimodal, diagonal sequence at later times. This includes the emergence of a sharp quiescent ridge in the upper-left region of the plane, corresponding to old, low-activity systems, which is absent at high redshift.
\textbf{AGN-dominated systems:} In contrast, these systems do not develop the bimodal structure or quiescent ridge. Instead, they remain concentrated along the activity axis (PC1), indicating a sustained phase of active growth rather than evolution toward quiescence.
\textbf{Global population:} The combined sample captures the full range of galaxy properties, with actively growing systems (young, high $f_{\mathrm{AGN}}$) occupying the lower region of the plane, and passive systems (old, low $f_{\mathrm{AGN}}$) located in the upper region.

Our analysis indicates that AGN luminosity is primarily governed by the host galaxy's current cold-gas supply, rather than its integrated history alone, with the instantaneous star-forming state serving as an observational tracer of this gas reservoir. Powerful AGNs are preferentially found in younger, gas-rich systems, a trend that becomes increasingly pronounced toward higher redshifts, where black hole growth and star formation appear to be tightly synchronized, consistent with both processes being fueled by a common gas supply.

In this work, the population of AGN-dominated systems evolves more strongly with cosmic time, this also includes selection effects. Their number decreases from 162 systems at early epochs to 31 in the most recent Universe, representing a significant decline. In contrast, host-dominated systems remain comparatively stable, with population sizes ranging from approximately 530 to 630 galaxies across all epochs.

In conclusion, the analysis of 2,709 galaxies spanning $\sim$13 Gyr of cosmic time reveals that galaxy evolution is fundamentally governed by two principal components: a scale and activity axis (PC1, $\sim$59\% of the variance) and a stellar maturation axis (PC2, $\sim$34\%). Across all redshifts, AGN luminosity ($L_{\mathrm{AGN}}$) is strongly correlated with star formation rates, young stellar and gas mass, indicating that black hole growth traces the active, gas-rich phase of galaxy evolution.  The evolution from a tight coupling between AGN activity and star formation at early times to a stronger association with dust emission at later epochs reflects the progressive maturation of the galaxy population from rapid growth phases to the diverse structures observed in the present-day Universe.

\section*{acknowledgments}
We thank our anonymous referee for their positive feedback and useful comments. This work is based on observations made with the NASA/E-SA/CSA JWST. The data were obtained from the Mikulski Archive for Space Telescopes at the Space Telescope Science Institute, operated by the Association of Universities for  Research in Astronomy, Inc., under NASA contract NAS5-03127 for JWST. 
T.G. acknowledges the support of the National Science and Technology Council of Taiwan through grants 114-2112-M-007 -026, 113-2112-M-007 -006, 113 -2927-I-007 -501, and 114-2123-M-001 -002. T.H. acknowledges the support of the National Science and Technology Council of Taiwan through grants 110-2112-M-005 -013 -MY3, 110-2112-M-007 -034 and 112-2123-M-001 -004. The authors used ChatGPT (OpenAI) during the manuscript writing process for language refinement and text drafting assistance. All content was reviewed and edited by the authors, who take full responsibility for the final version of the manuscript.

{\it Software:}
  CIGALE \citep{Boquien2019},
  EAZY \citep{Brammer2008},
  Astropy \citep{Astropy2013, Astropy2018, Astropy2022},
  Numpy \citep{Harris2020},
  SciPy \citep{Virtanen2020},
  pandas \citep{McKinney2010},
  scikit-learn \citep{Pedregosa2011},
  Matplotlib \citep{Hunter2007}.

\section*{Author Contributions}
P.J. came up with the initial research concept, led the project, performed the analysis, and wrote the manuscript. T.G., C.T.L. and A.R.B. helped obtain and process data. T.H. assisted in volume limited sample analysis. All the co-authors assisted with the interpretation of results, discussion, and manuscript review. All authors approved the final version of the manuscript.

\bibliography{reference}{}
\bibliographystyle{aasjournalv7}

\appendix
\section{Correlations}

\begin{table*}
\centering
\resizebox{\textwidth}{!}{%
\begin{tabular}{llccc}
\hline
Parameter & Redshift bin & Total & AGN dominated & Host dominated\\
\hline
${\rm SFR}_{10\,{\rm Myr}}$ & $0.00-1.00$ & \begin{tabular}{l}$r=0.69\pm0.07,\ P=1.2e-95$\\$m=0.90\pm0.05,\ c=34.12\pm0.05$\end{tabular} & \begin{tabular}{l}$r=0.84\pm0.04,\ P=3.7e-09$\\$m=0.89\pm0.08,\ c=35.52\pm0.07$\end{tabular} & \begin{tabular}{l}$r=0.73\pm0.08,\ P=3.4e-108$\\$m=0.94\pm0.06,\ c=34.08\pm0.06$\end{tabular} \\
 & $1.00-1.75$ & \begin{tabular}{l}$r=0.59\pm0.02,\ P=1.4e-63$\\$m=0.73\pm0.04,\ c=34.42\pm0.02$\end{tabular} & \begin{tabular}{l}$r=0.86\pm0.04,\ P=4.5e-09$\\$m=0.86\pm0.08,\ c=35.88\pm0.06$\end{tabular} & \begin{tabular}{l}$r=0.66\pm0.03,\ P=1.5e-81$\\$m=0.77\pm0.05,\ c=34.35\pm0.02$\end{tabular} \\
 & $1.75-2.75$ & \begin{tabular}{l}$r=0.60\pm0.03,\ P=9.6e-69$\\$m=0.67\pm0.04,\ c=34.75\pm0.03$\end{tabular} & \begin{tabular}{l}$r=0.90\pm0.03,\ P=5.0e-21$\\$m=1.07\pm0.08,\ c=35.65\pm0.08$\end{tabular} & \begin{tabular}{l}$r=0.68\pm0.03,\ P=1.7e-87$\\$m=0.67\pm0.05,\ c=34.66\pm0.04$\end{tabular} \\
 & $2.75-6.00$ & \begin{tabular}{l}$r=0.49\pm0.03,\ P=1.3e-43$\\$m=0.60\pm0.04,\ c=35.37\pm0.05$\end{tabular} & \begin{tabular}{l}$r=0.80\pm0.03,\ P=6.8e-37$\\$m=0.98\pm0.06,\ c=35.83\pm0.07$\end{tabular} & \begin{tabular}{l}$r=0.64\pm0.03,\ P=3.0e-62$\\$m=0.64\pm0.05,\ c=35.12\pm0.06$\end{tabular} \\
\hline
SFH Age & $0.00-1.00$ & \begin{tabular}{l}$r=0.16\pm0.04,\ P=2.7e-05$\\$m=0.00\pm0.00,\ c=33.08\pm0.19$\end{tabular} & \begin{tabular}{l}$r=-0.12\pm0.08,\ P=5.2e-01$\\$m=-0.00\pm0.00,\ c=34.81\pm0.13$\end{tabular} & \begin{tabular}{l}$r=0.21\pm0.05,\ P=9.2e-08$\\$m=0.00\pm0.00,\ c=32.88\pm0.21$\end{tabular} \\
 & $1.00-1.75$ & \begin{tabular}{l}$r=0.37\pm0.03,\ P=2.5e-22$\\$m=0.00\pm0.00,\ c=33.98\pm0.05$\end{tabular} & \begin{tabular}{l}$r=0.28\pm0.08,\ P=1.5e-01$\\$m=0.00\pm0.00,\ c=35.40\pm0.15$\end{tabular} & \begin{tabular}{l}$r=0.40\pm0.03,\ P=1.1e-25$\\$m=0.00\pm0.00,\ c=33.93\pm0.05$\end{tabular} \\
 & $1.75-2.75$ & \begin{tabular}{l}$r=0.05\pm0.03,\ P=2.2e-01$\\$m=0.00\pm0.00,\ c=35.07\pm0.06$\end{tabular} & \begin{tabular}{l}$r=0.30\pm0.07,\ P=2.2e-02$\\$m=0.00\pm0.00,\ c=35.60\pm0.12$\end{tabular} & \begin{tabular}{l}$r=0.12\pm0.04,\ P=1.7e-03$\\$m=0.00\pm0.00,\ c=34.90\pm0.06$\end{tabular} \\
 & $2.75-6.00$ & \begin{tabular}{l}$r=-0.14\pm0.03,\ P=2.3e-04$\\$m=-0.00\pm0.00,\ c=36.08\pm0.05$\end{tabular} & \begin{tabular}{l}$r=0.04\pm0.06,\ P=6.1e-01$\\$m=0.00\pm0.00,\ c=36.42\pm0.09$\end{tabular} & \begin{tabular}{l}$r=-0.06\pm0.03,\ P=1.6e-01$\\$m=-0.00\pm0.00,\ c=35.78\pm0.06$\end{tabular} \\
\hline
$M_\star/M_\odot$ & $0.00-1.00$ & \begin{tabular}{l}$r=0.61\pm0.08,\ P=5.5e-70$\\$m=0.71\pm0.11,\ c=27.11\pm0.99$\end{tabular} & \begin{tabular}{l}$r=0.72\pm0.02,\ P=4.1e-06$\\$m=0.73\pm0.02,\ c=28.63\pm0.20$\end{tabular} & \begin{tabular}{l}$r=0.69\pm0.10,\ P=2.2e-91$\\$m=0.81\pm0.13,\ c=26.17\pm1.14$\end{tabular} \\
 & $1.00-1.75$ & \begin{tabular}{l}$r=0.63\pm0.02,\ P=2.7e-75$\\$m=0.58\pm0.03,\ c=29.06\pm0.27$\end{tabular} & \begin{tabular}{l}$r=0.79\pm0.03,\ P=6.8e-07$\\$m=0.67\pm0.06,\ c=29.76\pm0.62$\end{tabular} & \begin{tabular}{l}$r=0.72\pm0.02,\ P=6.4e-102$\\$m=0.61\pm0.03,\ c=28.76\pm0.30$\end{tabular} \\
 & $1.75-2.75$ & \begin{tabular}{l}$r=0.46\pm0.03,\ P=4.8e-37$\\$m=0.42\pm0.04,\ c=31.22\pm0.36$\end{tabular} & \begin{tabular}{l}$r=0.86\pm0.03,\ P=9.7e-18$\\$m=0.88\pm0.06,\ c=28.29\pm0.59$\end{tabular} & \begin{tabular}{l}$r=0.57\pm0.04,\ P=2.5e-55$\\$m=0.46\pm0.05,\ c=30.75\pm0.43$\end{tabular} \\
 & $2.75-6.00$ & \begin{tabular}{l}$r=0.34\pm0.02,\ P=1.6e-20$\\$m=0.36\pm0.04,\ c=32.49\pm0.35$\end{tabular} & \begin{tabular}{l}$r=0.74\pm0.03,\ P=6.6e-30$\\$m=0.83\pm0.06,\ c=29.02\pm0.54$\end{tabular} & \begin{tabular}{l}$r=0.51\pm0.03,\ P=2.3e-36$\\$m=0.46\pm0.05,\ c=31.40\pm0.46$\end{tabular} \\
\hline
${\rm Age}_\star$ & $0.00-1.00$ & \begin{tabular}{l}$r=0.08\pm0.05,\ P=3.3e-02$\\$m=0.00\pm0.00,\ c=33.40\pm0.14$\end{tabular} & \begin{tabular}{l}$r=-0.05\pm0.08,\ P=7.9e-01$\\$m=-0.00\pm0.00,\ c=34.69\pm0.10$\end{tabular} & \begin{tabular}{l}$r=0.11\pm0.05,\ P=4.2e-03$\\$m=0.00\pm0.00,\ c=33.29\pm0.15$\end{tabular} \\
 & $1.00-1.75$ & \begin{tabular}{l}$r=0.36\pm0.03,\ P=9.2e-22$\\$m=0.00\pm0.00,\ c=34.13\pm0.04$\end{tabular} & \begin{tabular}{l}$r=0.23\pm0.07,\ P=2.4e-01$\\$m=0.00\pm0.00,\ c=35.53\pm0.12$\end{tabular} & \begin{tabular}{l}$r=0.39\pm0.03,\ P=5.2e-24$\\$m=0.00\pm0.00,\ c=34.09\pm0.04$\end{tabular} \\
 & $1.75-2.75$ & \begin{tabular}{l}$r=0.04\pm0.03,\ P=3.3e-01$\\$m=0.00\pm0.00,\ c=35.09\pm0.05$\end{tabular} & \begin{tabular}{l}$r=0.32\pm0.07,\ P=1.5e-02$\\$m=0.00\pm0.00,\ c=35.65\pm0.11$\end{tabular} & \begin{tabular}{l}$r=0.10\pm0.04,\ P=1.2e-02$\\$m=0.00\pm0.00,\ c=34.95\pm0.05$\end{tabular} \\
 & $2.75-6.00$ & \begin{tabular}{l}$r=-0.10\pm0.03,\ P=9.1e-03$\\$m=-0.00\pm0.00,\ c=35.99\pm0.04$\end{tabular} & \begin{tabular}{l}$r=0.09\pm0.05,\ P=2.5e-01$\\$m=0.00\pm0.00,\ c=36.37\pm0.07$\end{tabular} & \begin{tabular}{l}$r=-0.05\pm0.03,\ P=2.6e-01$\\$m=-0.00\pm0.00,\ c=35.75\pm0.05$\end{tabular} \\
\hline
$L_{\rm dust}$ & $0.00-1.00$ & \begin{tabular}{l}$r=0.82\pm0.09,\ P=3.2e-167$\\$m=0.83\pm0.03,\ c=3.98\pm1.16$\end{tabular} & \begin{tabular}{l}$r=0.83\pm0.05,\ P=9.5e-09$\\$m=0.70\pm0.06,\ c=10.16\pm2.04$\end{tabular} & \begin{tabular}{l}$r=0.87\pm0.09,\ P=1.7e-193$\\$m=0.86\pm0.03,\ c=2.89\pm1.16$\end{tabular} \\
 & $1.00-1.75$ & \begin{tabular}{l}$r=0.72\pm0.02,\ P=2.9e-104$\\$m=0.65\pm0.03,\ c=10.94\pm1.15$\end{tabular} & \begin{tabular}{l}$r=0.83\pm0.05,\ P=3.9e-08$\\$m=0.71\pm0.08,\ c=10.15\pm3.09$\end{tabular} & \begin{tabular}{l}$r=0.79\pm0.03,\ P=9.7e-137$\\$m=0.66\pm0.03,\ c=10.58\pm1.13$\end{tabular} \\
 & $1.75-2.75$ & \begin{tabular}{l}$r=0.59\pm0.03,\ P=3.0e-65$\\$m=0.50\pm0.03,\ c=16.70\pm1.21$\end{tabular} & \begin{tabular}{l}$r=0.90\pm0.03,\ P=3.7e-21$\\$m=0.85\pm0.06,\ c=5.03\pm2.27$\end{tabular} & \begin{tabular}{l}$r=0.68\pm0.04,\ P=1.6e-88$\\$m=0.51\pm0.04,\ c=16.38\pm1.39$\end{tabular} \\
 & $2.75-6.00$ & \begin{tabular}{l}$r=0.49\pm0.02,\ P=1.2e-42$\\$m=0.46\pm0.03,\ c=18.90\pm1.24$\end{tabular} & \begin{tabular}{l}$r=0.84\pm0.03,\ P=3.6e-44$\\$m=0.77\pm0.05,\ c=8.11\pm1.93$\end{tabular} & \begin{tabular}{l}$r=0.66\pm0.03,\ P=1.0e-66$\\$m=0.53\pm0.04,\ c=16.08\pm1.66$\end{tabular} \\
\hline

\end{tabular}
}
\caption{
Redshift-dependent Pearson correlation coefficients and
linear regression parameters for the relation between
AGN luminosity and host galaxy properties.
The linear relation is defined as
$y=mx+c$.
}
\label{tab:corr}
\end{table*}

\section{Principal Component Analysis of Volume-Limited Subsamples}
\label{sec:appendix}
To test whether the observed evolutionary trends are robust against luminosity-dependent selection effects, we performed PCA on three independent volume-limited subsamples spanning different ranges of AGN luminosity and lookback time. By isolating galaxies within restricted luminosity intervals, this approach enables a more reliable comparison of galaxy and black hole co-evolution across cosmic time.

\begin{figure}
    \centering
    \includegraphics[width=0.49\textwidth]{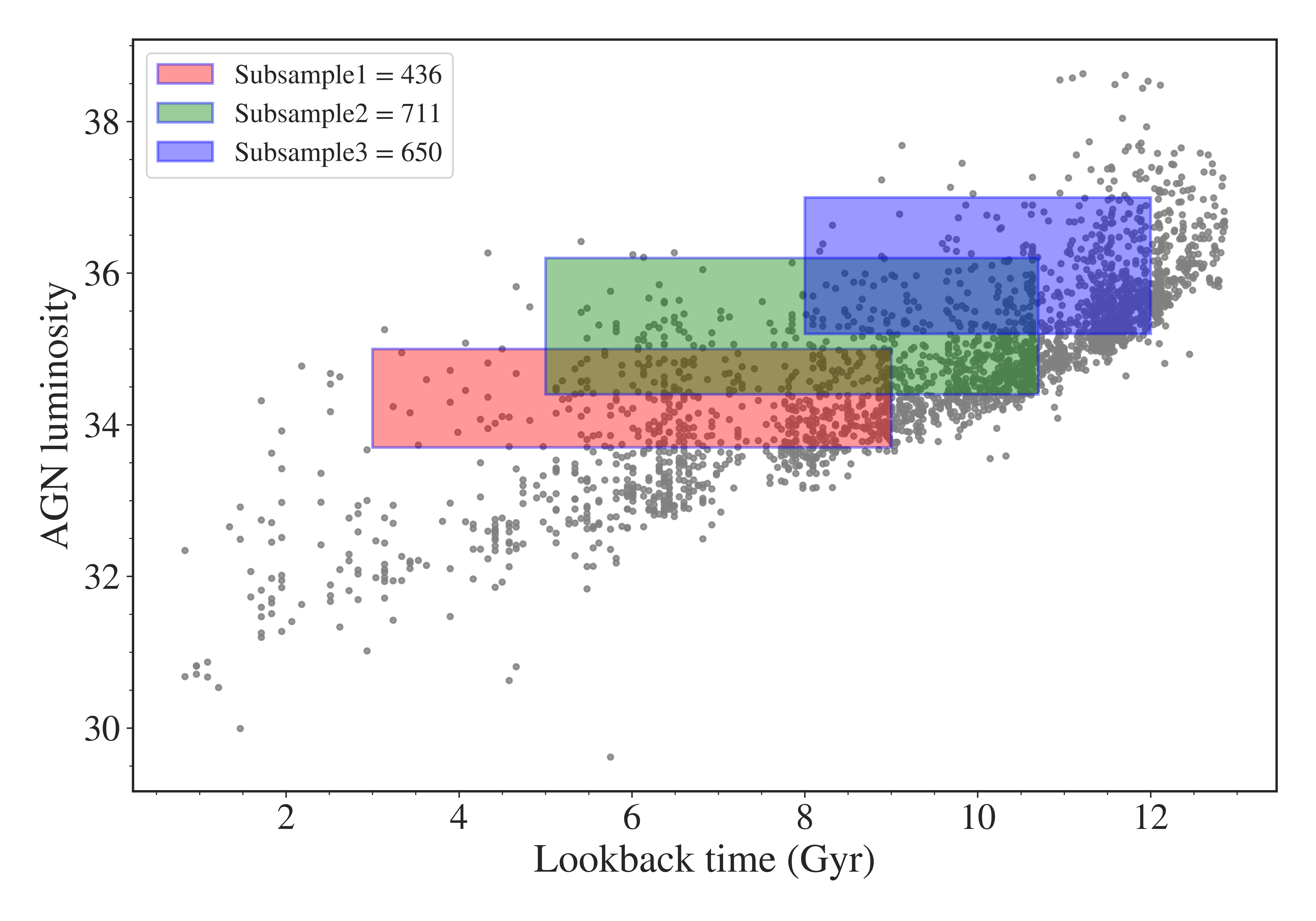}
       \caption{Distribution of AGN bolometric luminosity as a function of lookback time for the full galaxy sample (grey points). The shaded regions define the three volume-limited PCA subsamples used to minimize luminosity-dependent selection effects: Subsample~1 (red; $N=436$) probes lower-luminosity AGN at recent epochs, Subsample~2 (green; $N=711$) samples intermediate luminosities at intermediate lookback times, and Subsample~3 (blue; $N=650$) isolates the most luminous AGN at high redshift. These selections enable consistent evolutionary comparisons within fixed luminosity regimes across cosmic time.
}
    \label{fig:subsample}
\end{figure}

\begin{figure*}
    \centering
    \includegraphics[width=0.49\textwidth]{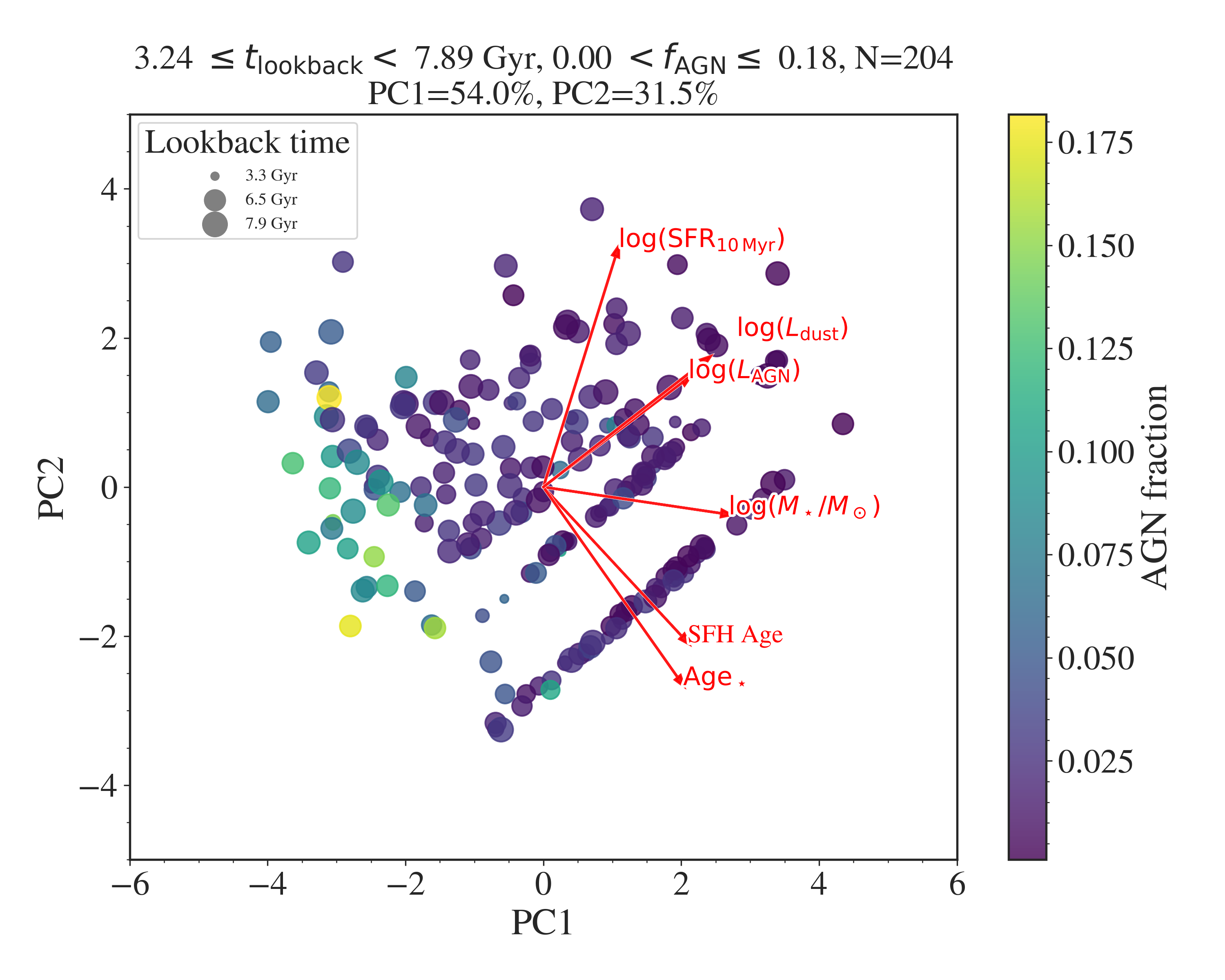}
       \includegraphics[width=0.49\textwidth]{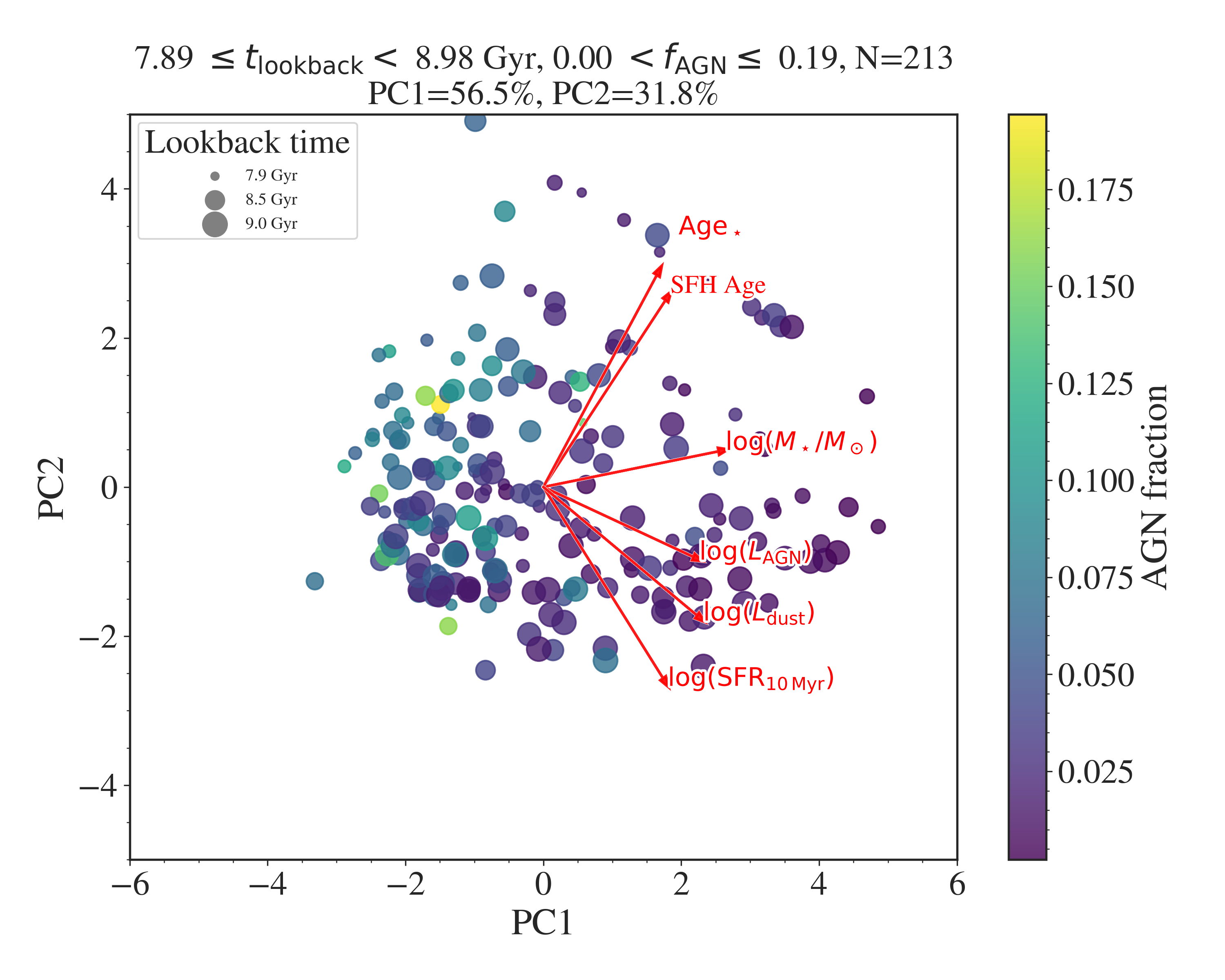}
       \caption{PCA projections for Subsample~1, tracing the evolution of lower-luminosity AGN hosts across two lookback-time intervals: recent ($3.2$--$7.9$ Gyr; left) and early ($7.9$--$9.0$ Gyr; right).}
    \label{fig:subsample1}
\end{figure*}

\begin{figure*}
    \centering
    \includegraphics[width=0.49\textwidth]{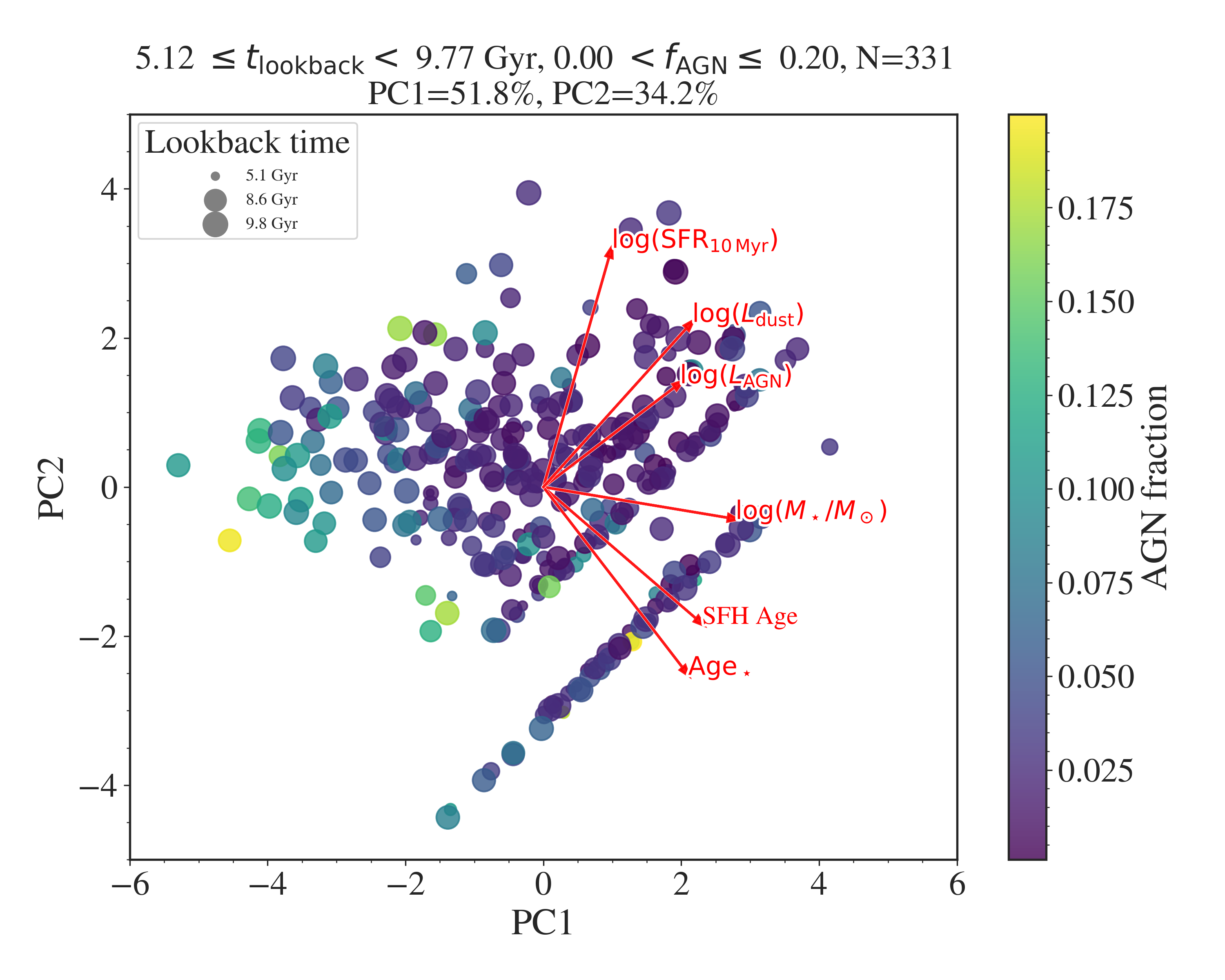}
       \includegraphics[width=0.49\textwidth]{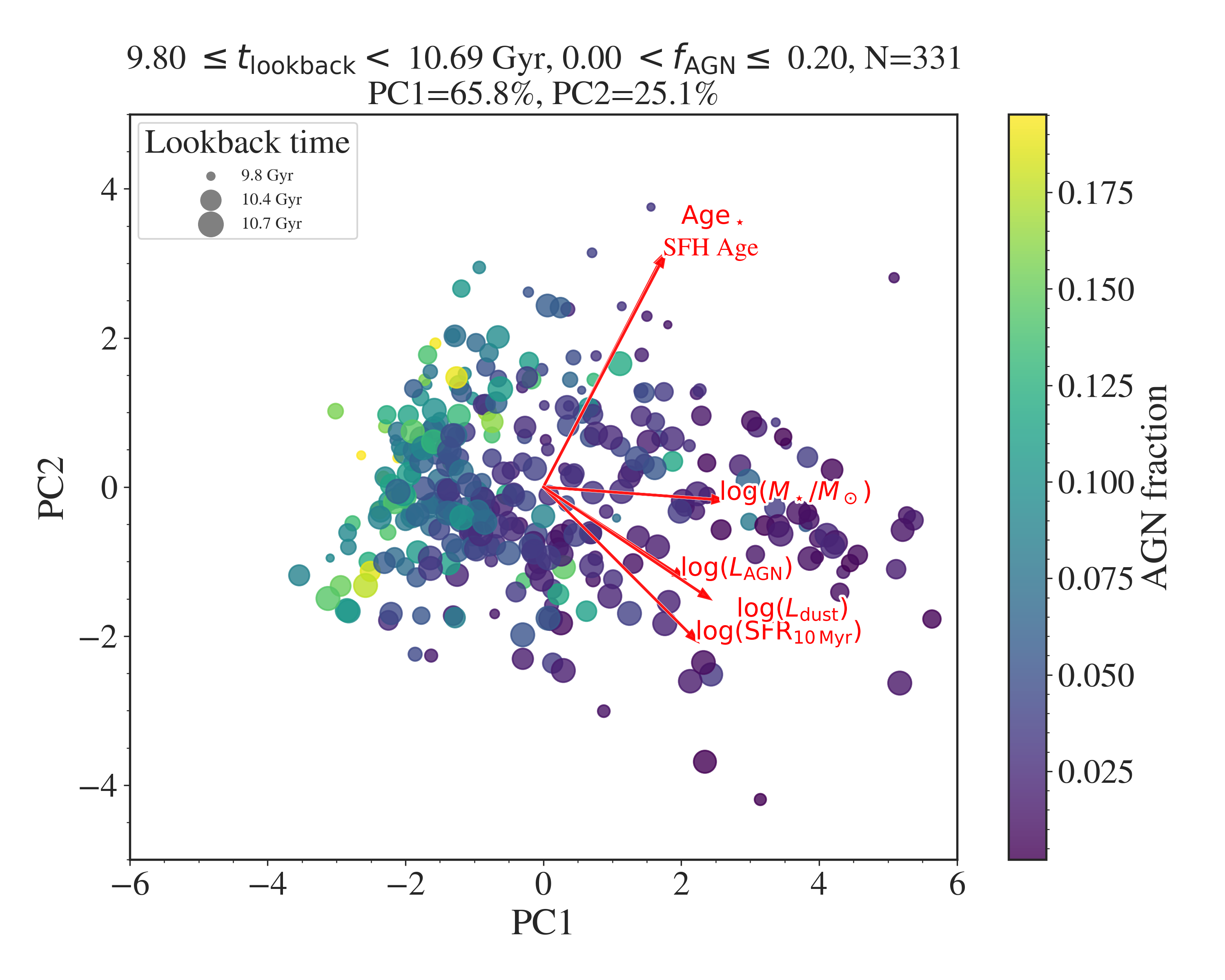}
       \caption{Same as Figure.~\ref{fig:subsample1} but for Subsample~2.}
    \label{fig:subsample2}
\end{figure*}

\begin{figure*}
    \centering
    \includegraphics[width=0.49\textwidth]{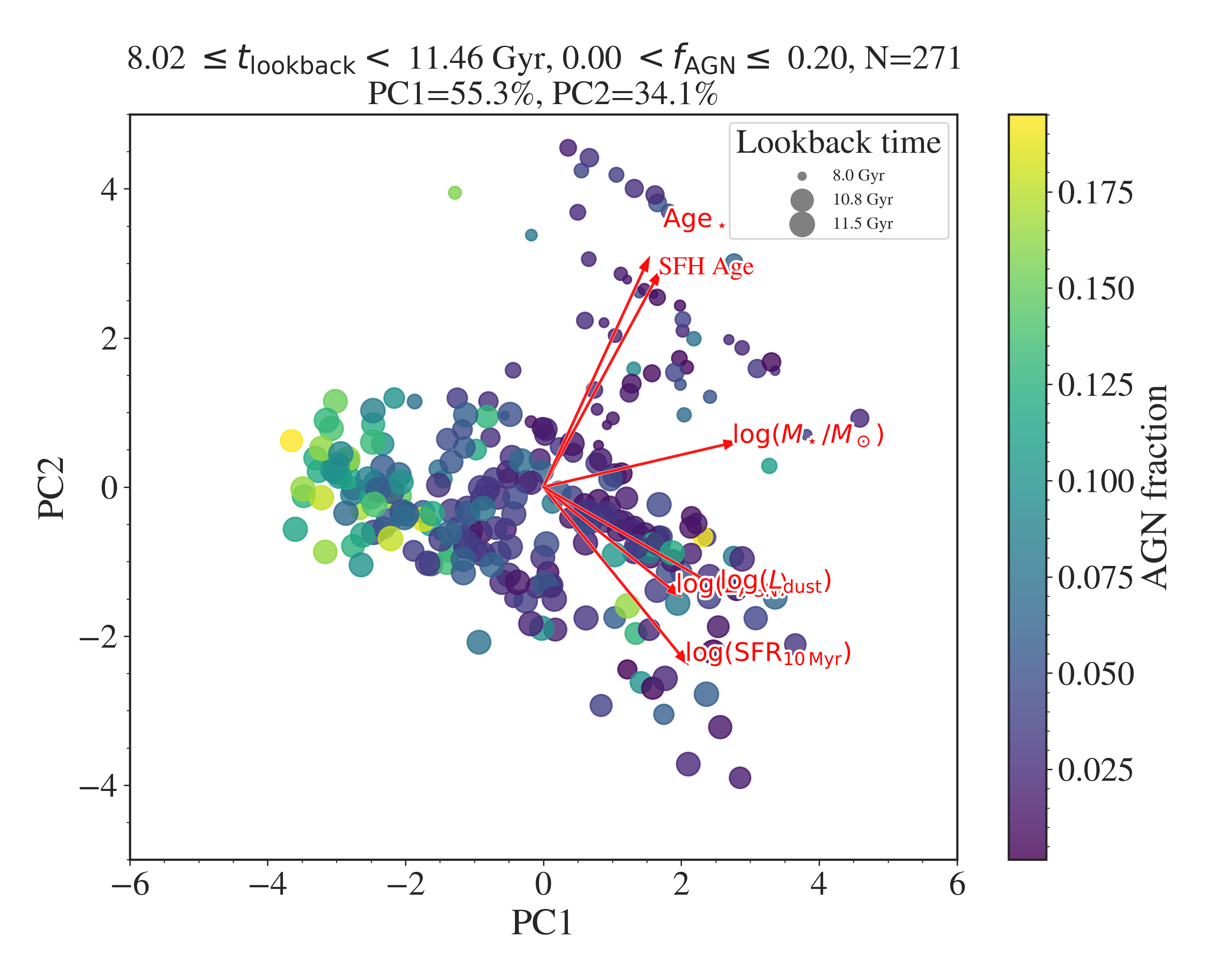}
       \includegraphics[width=0.49\textwidth]{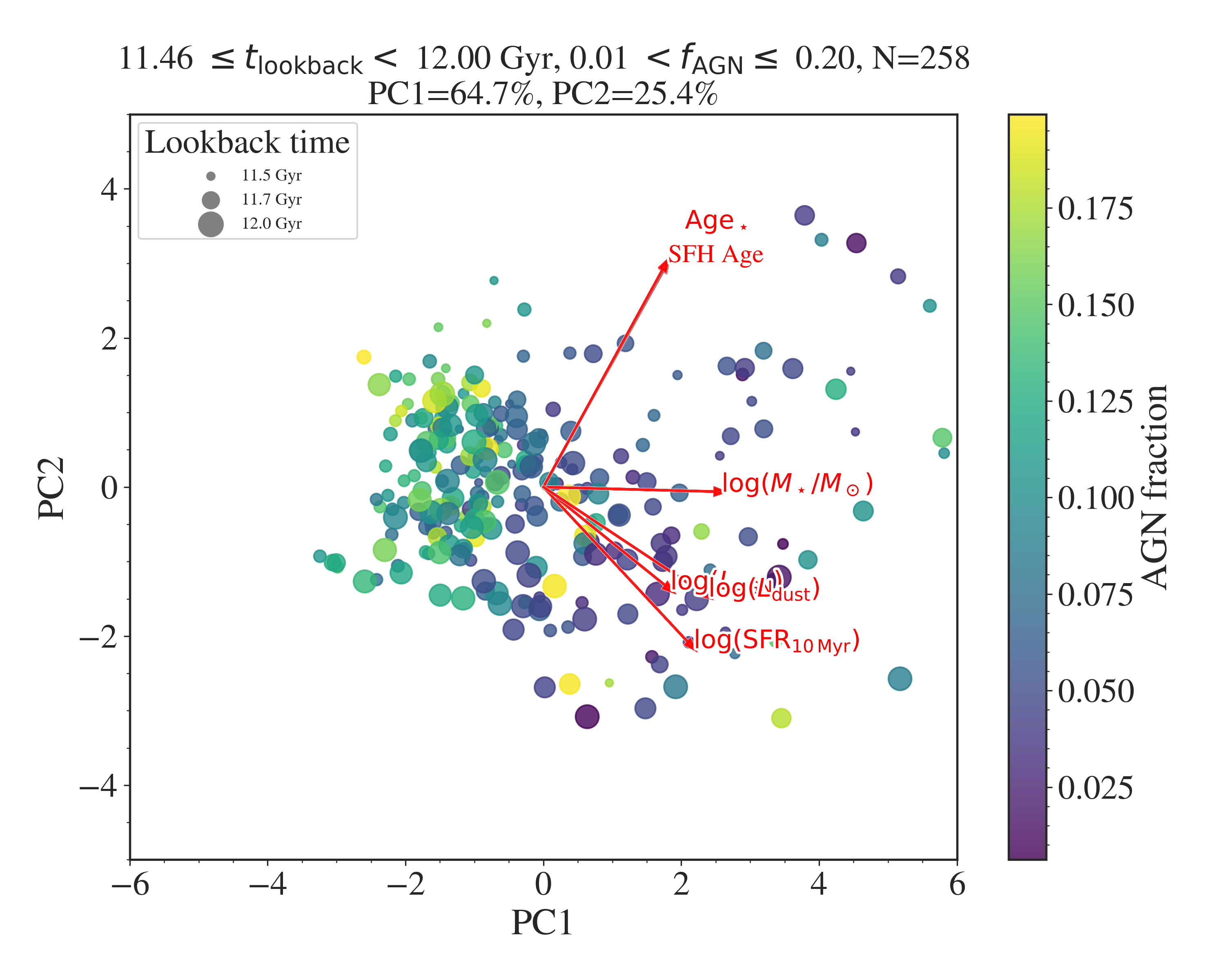}
       \caption{Same as Figure.~\ref{fig:subsample1} but for Subsample~3.}
    \label{fig:subsample3}
\end{figure*}

The three subsamples were selected according to $L_{\mathrm{AGN}}$ and lookback time distributions, as illustrated in  Figure.~\ref{fig:subsample}. The adopted ranges are summarized below:

\begin{itemize}
    \item \textbf{Subsample 1 (Red):} 436 galaxies with relatively low AGN luminosities,
    \[
    33.7 \lesssim \log L_{\mathrm{AGN}} \lesssim 35,
    \]
    spanning lookback times of approximately $3$--$9$ Gyr (Figure.~\ref{fig:subsample1}).

    \item \textbf{Subsample 2 (Green):} 711 galaxies with intermediate AGN luminosities,
    \[
    34.4 \lesssim \log L_{\mathrm{AGN}} \lesssim 36.2,
    \]
    spanning lookback times of approximately $5$--$10.7$ Gyr  (Figure.~\ref{fig:subsample2}).

    \item \textbf{Subsample 3 (Blue):} 650 galaxies with the highest AGN luminosities,
    \[
    35.2 \lesssim \log L_{\mathrm{AGN}} \lesssim 37,
    \]
    spanning lookback times of approximately $8$--$12$ Gyr  (Figure.~\ref{fig:subsample3}).
\end{itemize}

Each subsample was further divided into two lookback-time bins in order to examine temporal evolution within a fixed luminosity regime. We do not divide the subsamples in host and AGN dominated samples due to limited sample size.

The higher lookback-time bins of all three subsamples reveal a remarkably coherent PCA structure indicative of synchronized galaxy and black hole growth in the early Universe. Consistent with previous results, in these epochs, the loading vectors associated with AGN luminosity, recent SFR, and dust luminosity are closely aligned along the positive PC1 axis, suggesting that black hole accretion and star formation were tightly coupled processes, both fuelled by a common reservoir of cold gas in gas-rich galaxies.

At lower lookback times, the PCA structure becomes increasingly complex. The activity-related vectors begin to separate or ``fan out,'' producing larger angular offsets between stellar mass AGN luminosity, and star formation indicators. This decoupling implies that high stellar mass no longer corresponds uniquely to elevated AGN or star-forming activity, reflecting the emergence of more diverse evolutionary pathways among galaxies.

A particularly notable feature of the lower-redshift PCA distributions is the appearance of a diagonal sequence of low-activity, older galaxies occupying the upper-left region of the PCA plane. This structure, most prominent in Subsample~1 ($3.2$--$7.9$ Gyr) and Subsample~2 ($5.1$--$9.8$ Gyr), forms a well-defined ``quiescent ridge'' associated with passive or quenched systems. 

Despite the differing luminosity regimes probed by the three subsamples, the overall PCA architecture remains remarkably stable. Across all subsamples, PC1 consistently traces global activity and scale, accounting for approximately $50$--$64\%$ of the total variance. The dominant contributors to this component are stellar mass, star formation rate, dust luminosity, and AGN luminosity.

The second principal component (PC2), accounting for roughly $24$--$32\%$ of the variance, is primarily associated with stellar population age and star formation history (SFH) age, thereby tracing evolutionary maturity.

In all subsamples, galaxies with the highest AGN fractions , highlighted in yellow in the PCA projections, systematically occupy the high-activity and low-age regions of parameter space. These systems are frequently positioned opposite the stellar mass vector, reinforcing the interpretation that fractional AGN dominance is preferentially enhanced in lower-mass galaxies.

The PCA analysis of the volume-limited subsamples similar to host dominated analysis (the total sample is dominated by host-dominated sample)  supports a coherent evolutionary picture in which the Universe transitions from an early phase of tightly synchronized, gas-driven growth to a later phase characterized by increasing demographic diversity and partial decoupling between black hole accretion and star formation. While the primary drivers of variance remain stable across luminosity regimes, the changing geometric structure of the PCA space reveals the progressive emergence of quiescent populations and heterogeneous evolutionary pathways over cosmic time.

\end{document}